\documentclass[11pt,a4paper,twoside,fleqn]{article}
\usepackage{color,graphicx,lscape,here,geometry,setspace,multirow,enumitem}
\usepackage[dvipsnames]{xcolor}
\usepackage{graphicx}
\usepackage{natbib}
\usepackage{authblk}
\usepackage{silence}
\usepackage{helvet}

\usepackage{rotating}
\usepackage{multirow}

\usepackage{booktabs}
\usepackage{diagbox}
\usepackage{color,colortbl}
\usepackage{setspace}
\usepackage{algorithm2e}
\usepackage{enumitem}
\usepackage{amsfonts}
\usepackage{tikz}

\definecolor{nblue}{HTML}{000660}
\usepackage[colorlinks=true,urlcolor=nblue,linkcolor=nblue,citecolor=nblue]{hyperref}
\usepackage{bigstrut}

\definecolor{colRWSV}{HTML}{999999}
\definecolor{colUCSV}{HTML}{d7af19}
\definecolor{colUCSVM}{HTML}{2c7bb6}
\definecolor{colQRUC}{HTML}{d7191c}

\newcommand{\colBench}{\cellcolor{gray!10}}
\newcommand{\colBest}{\cellcolor{gray!30}}

\usepackage{tabularx}
\usepackage{threeparttable}
\usepackage{dcolumn}
\newcolumntype{d}[1]{D{.}{.}{#1}}

\usepackage{array}
\newcolumntype{C}[1]{>{\centering\arraybackslash}p{#1}}

\usepackage{etoolbox} 
\makeatletter
\patchcmd{\BR@backref}{\newblock}{\newblock[}{}{}
\patchcmd{\BR@backref}{\par}{]\par}{}{}
\makeatother

\usepackage{pdflscape}
\usepackage{afterpage}

\usepackage{longtable}
\usepackage{array}

\usepackage{comment}
\usepackage{mathtools}
\usepackage[hang,flushmargin]{footmisc} 
\usepackage[hang]{footmisc}
\usepackage[british]{babel}

\pdfoutput=1
\usepackage{float}
\restylefloat{table}

\usepackage[title,titletoc]{appendix}
\makeatletter

\renewenvironment{appendices}{%
    \begin{oldappendices}%
    \renewcommand{\thefigure}{\ifnum \c@section>\z@ \thesection.\fi\@arabic\c@figure}%
    \@addtoreset{figure}{section}%
    \renewcommand{\thetable}{\ifnum \c@section>\z@ \thesection.\fi\@arabic\c@table}%
    \@addtoreset{table}{section}}{%
    \end{oldappendices}%
}\makeatother

\usepackage{titlesec} 
\titleformat{\section}[block]{\sffamily\large}{\thesection. }{0em}{\MakeUppercase} 
\titleformat{\subsection}[block]{\sffamily\large}{\thesubsection. }{0em}{\itshape} 
\titleformat{\subsubsection}[block]{\sffamily\large}{}{0em}{\itshape} 

\let\natbibcitet\citet
\renewcommand\citet{\bibpunct{(}{)}{,}{a}{,}{,}\natbibcitet}
\let\natbibcitep\citep
\renewcommand\citep{\bibpunct{(}{)}{;}{a}{,}{;}\natbibcitep}
\newcommand{\bi}{\begin{itemize}}
\newcommand{\ei}{\end{itemize}}
\newcommand{\be}{\begin{equation}}
\newcommand{\ee}{\end{equation}}
\defcitealias{ieo14}{IEO, 2014}

\long\def\symbolfootnote[#1]#2{\begingroup%
\def\thefootnote{\fnsymbol{footnote}}\footnote[#1]{#2}\endgroup}

\makeatletter
\def\ubar#1{\underline{\sbox\tw@{$#1$}\dp\tw@\z@\box\tw@}}
\def\obar#1{\overline{\sbox\tw@{$#1$}\dp\tw@\z@\box\tw@}}
\makeatother

\usepackage{bm}
\usepackage{caption}
\usepackage{subcaption}

\makeatletter\let\p@subfigure\thefigure\makeatother

\floatstyle{plaintop}
\restylefloat{table}

\usepackage{fancyhdr} 
\usepackage{cleveref}
\crefname{chapter}{Chapter}{Chapters}
\crefname{section}{Section}{Sections}
\crefname{subsection}{Section}{Sections}
\crefname{subsubsection}{Section}{Sections}
\crefname{figure}{Figure}{Figures}
\crefname{table}{Table}{Tables}
\crefname{equation}{Equation}{Equations}
\crefname{appendix}{Appendix}{Appendices}
\crefname{appendices}{Appendix}{Appendices}

\crefname{appsec}{Appendix}{Appendices}

\usepackage{catoptions}
\makeatletter

\def\Autoref#1{%
  \begingroup
  \edef\reserved@a{\cpttrimspaces{#1}}%
  \ifcsndefTF{r@#1}{%
    \xaftercsname{\expandafter\testreftype\@fourthoffive}
      {r@\reserved@a}.\\{#1}%
  }{%
    \ref{#1}%
  }%
  \endgroup
}
\def\testreftype#1.#2\\#3{%
  \ifcsndefTF{#1autorefname}{%
    \def\reserved@a##1##2\@nil{%
      \uppercase{\def\ref@name{##1}}%
      \csn@edef{#1autorefname}{\ref@name##2}%
      \autoref{#3}%
    }%
    \reserved@a#1\@nil
  }{%
    \autoref{#3}%
  }%
}
\makeatother

\title{\sffamily\LARGE{\textbf{Modeling tail risks of inflation using unobserved component quantile regressions}}}
\author{\large{
\uppercase{Michael Pfarrhofer}}\thanks{
\noindent Department of Economics, University of Salzburg. \textit{Address}: M\"{o}nchsberg 2a, 5020 Salzburg, Austria. \textit{Email}: \href{mailto:michael.pfarrhofer@plus.ac.at}{michael.pfarrhofer@plus.ac.at}. This paper was previously circulated as ``Tail forecasts of inflation using time-varying parameter quantile regressions'' (arXiv:2103.03632). I gratefully acknowledge valuable comments and suggestions by Niko Hauzenberger, Paul Hofmarcher, Florian Huber, Gary Koop, James Mitchell and Anna Stelzer, and financial support by the Austrian Science Fund (FWF, grant no. ZK 35). Codes and replication files are available at \href{https://github.com/mpfarrho/tvp-qr}{github.com/mpfarrho/tvp-qr}.}
\\\vspace*{-0.5em}
\textit{University of Salzburg}}
\date{}


\def\equationautorefname~#1\null{%
  Eq.~(#1)\null
}
\def\equationautorefname~#1\null{
Eq.~(#1)\null
}

\usepackage[utf8, latin1]{inputenc}                 

\begin{document}
\maketitle\thispagestyle{empty}\normalsize\vspace*{-2em}\small

\begin{center}
\begin{minipage}{0.8\textwidth}
\noindent\small This paper proposes methods for Bayesian inference in time-varying parameter (TVP) quantile regression (QR) models featuring conditional heteroskedasticity. I use data augmentation schemes to render the model conditionally Gaussian and develop an efficient Gibbs sampling algorithm. Regularization of the high-dimensional parameter space is achieved via flexible dynamic shrinkage priors. A simple version of TVP-QR based on an unobserved component model is applied to dynamically trace the quantiles of the distribution of inflation in the United States, the United Kingdom and the euro area. In an out-of-sample forecast exercise, I find the proposed model to be competitive and perform particularly well for higher-order and tail forecasts. A detailed analysis of the resulting predictive distributions reveals that they are sometimes skewed and occasionally feature heavy tails.
\end{minipage}
\end{center}

\doublespacing\normalsize\renewcommand{\thepage}{\arabic{page}}\newpage

\section{Introduction}\label{sec:introduction}
Predictive inference is often concerned with producing point forecasts of the conditional mean of some economic or financial series. However, recent contributions to the forecasting literature stress to take into account higher-order moments of the predictive distribution. Tail risks in particular have been studied in the finance literature for some time, but interest in the tails of distributions of key macroeconomic variables is a comparatively recent phenomenon. In an influential paper, \citet{adrian2019vulnerable} find time-varying downside risks in the distribution of GDP growth conditional on economic and financial conditions. \citet{adams2021forecasting} use a similar framework to quantify risks to several other key macroeconomic variables, and come to the same conclusions: risks are time-varying, potentially asymmetric, and partly predictable.\footnote{Other recent papers that address tail risks of macroeconomic variables are \citet{manzan2015forecasting,giglio2016systemic,denicolo2017forecasting,kiley2018unemployment,galbraith2019asymmetry,caldara2020macroeconomic,carriero2020bcapturing,carriero2020anowcasting,plagborg2020growth,clark2021tail,clark2021investigating,delle2021modeling}.}

A popular approach to model tail risks is quantile regression (QR). QRs are designed to estimate the conditional quantiles of some endogenous variable, and have originally been proposed by \citet{koenker1978regression}. In this paper I add to the literature on QRs featuring time-varying parameters (TVPs) in a unified state space framework. TVPs have proven useful in addressing parameter change in macroeconomic and financial series in the context of structural inference, and often improve predictive accuracy when interest centers on forecasting.\footnote{Examples using TVP models for structural inference are \citet{primiceri2005time,mumtaz2018changing,paul2020time}. Recent forecasting applications featuring TVP models are, for instance, \citet{d2013macroeconomic,aastveit2017have,huber2020inducing,yousuf2020boosting}.} 

Several previous contributions consider varying coefficient QRs \citep[see][]{de2006time,kim2007quantile,wang2009quantile,oka2011estimating}; others rely on approximate or nonparametric approaches to account for structural breaks when estimating higher-order moments of the distribution of some series \citep[see][]{cai2008quantile,taddy2010bayesian,gerlach2011bayesian,chen2013semi,liu2016markov,wu2017nonparametric,gonccalves2020dynamic,lim2020sparse,griffin2020bayesian}. A recent related paper is \citet{korobilis2021time}, who study time-varying inflation risks conditional on a set of macroeconomic and financial indicators in the euro area using a variant of TVP-QR.

The conventional Bayesian QR originates from \citet{yu2001bayesian}, who establish a correspondence to classical methods by specifying the likelihood function as an asymmetric Laplace distribution. One shortcoming of this approach is that it complicates the setup of an efficient Gibbs sampling algorithm. This is due to the fact that conditional posterior distributions of the model parameters are available in well-known form only for specific prior choices. As a solution, \citet{kozumi2011gibbs} introduce auxiliary variables to approximate the asymmetric Laplace distribution, which renders the model conditionally Gaussian. This approximation is the point of departure for using state space methods within the class of QR models.

I show how to extend the Bayesian QR to feature TVPs using conventional methods for Gaussian state space models involving several reparameterizations and approximations to facilitate the conditional likelihood. The assumption of a constant scale parameter of the asymmetric Laplace likelihood is relaxed, allowing for conditional heteroskedasticity within quantiles. While QR is itself a method to account for heteroskedastic data features, this extension adds an additional layer of flexibility. In particular, it allows the model to decide whether parameter change is attributed to the conditional quantile or whether quantiles feature time-variation in their error component, a crucial aspect when incorporating TVPs to discriminate signals from noise \citep[see, e.g., the discussion in][]{sims2001comment}. Since quantile estimates are not guaranteed to be monotonic in the baseline version of the model, I consider an auxiliary Gaussian process regression to provide estimates of noncrossing quantiles. A special case of the Bayesian TVP-QR is applied to model the distribution and to produce tail forecasts of inflation.

Forecasts of the conditional mean of inflation, but also its dispersion and tail risks, are of crucial importance to policy makers in central banks and practitioners in the private sector. Several related approaches using QRs have been proposed to model the full distribution of inflation or varying degrees of persistence in specific quantiles in autoregressive frameworks conditional on a set of other macroeconomic indicators \citep[see, e.g.,][]{wolters2015changing,lopez2020inflation}. From a forecasting perspective, some papers suggest improvements in predictive accuracy by directly modeling conditional quantiles of inflation \citep[e.g.,][]{manzan2013macroeconomic,korobilis2017quantile,ghysels2018quantile,korobilis2021time}.

Most of the preceding literature on quantile models of inflation includes explanatory variables based on stylized models such as the Phillips curve and small information sets \citep[e.g.,][]{manzan2013macroeconomic,lopez2020inflation}, or model/variable selection approaches in higher-dimensional data environments \citep[e.g.,][]{korobilis2017quantile}. While the methods proposed in this paper apply to the general case featuring regressors, I consider a QR-version of a comparatively simple TVP model that has had great success in forecasting the conditional mean of inflation in the empirical application: the unobserved component (UC) model of \citet{stock2007has}. Variants and extensions of this model featuring several additional unobserved factors \citep[e.g.,][]{chan2013new,chan2018new,jarocinski2018inflation}, or the UC stochastic volatility in mean (UC-SVM) model proposed by \citet{chan2017stochastic} exhibit strong overall forecast performance for inflation.

These models mainly target the conditional mean of inflation. Probabilistic error distributions allow to compute measures of density forecast accuracy and quantiles as a by-product. In this paper, I shed light on the question whether it is beneficial to model the quantiles of inflation explicitly within the class of unobserved component models, and how the resulting shapes of the predictive distributions differ along key dimensions such as skewness or heavy tails. Special emphasis is put on investigating the performance of such approaches in different parts of the predictive distribution. A version of TVP-QR featuring only an intercept for each quantile is used to model inflation in the United States (US), United Kingdom (UK) and the euro area (EA). The applied model is an unobserved component quantile regression, subsequently labeled UCQR. I alleviate overfitting concerns that are common in TVP models via adopting dynamic shrinkage priors to regularize the parameter space.

Dynamic shrinkage is required when the underlying series or relation between variables exhibits a mixture of periods characterized by stable dynamics, abrupt structural breaks or moderate time-variation. I compare three distinct prior specifications. The first is a conventional time-invariant setup that disregards dynamic shrinkage. By contrast, there are two options to shrink dynamically. One may either impose shrinkage period-by-period, ruling out a persistent shrinkage process. This is the second variant I consider, based on the horseshoe global-local prior. The third is the dynamic horseshoe. Here, the degree of shrinkage over time is informed based on a persistent process. Tail observations occur infrequently by definition, and this property could be particularly useful in the case of UCQR.

I compare density and tail forecasts for three model specifications: the UC-SV model, the UC-SVM model and the UCQR model. The latter is estimated using several alternative specifications for imposing shrinkage on the TVPs to assess empirical properties of the dynamic shrinkage priors, with and without time-varying scale parameter, and post-processed in two ways to guarantee monotonic quantile estimates. The results suggest that variants of UCQR perform particularly well for higher-order and tail forecasts. Among the considered UCQR-models, there is some heterogeneity with respect to the considered economies. Conditional heteroskedasticity is required to improve predictive accuracy in the US and the EA, while homoskedasticity is sufficient in the UK. The dynamic horseshoe exhibits the most consistent performance across forecast metrics, horizons and economies. A detailed analysis of the resulting predictive distributions reveals that they are sometimes skewed and occasionally feature heavy tails.

The rest of the paper is structured as follows. Section \ref{sec:econometrics} presents a general version of the TVP-QR model and adopts priors for imposing shrinkage dynamically. Section \ref{sec:motivation} discusses details about UCQR as a special case of the general framework, and motivates this specification in light of stylized facts about inflation. Sections \ref{sec:insample} and \ref{sec:forecasts} apply the model to a study of inflation dynamics in the US, UK and EA, and contain in-sample results alongside predictive inference. Section \ref{sec:conclusions} offers concluding remarks. Appendices \ref{app:A:Sampling} and \ref{app:B:Results} contain additional details on Bayesian inference and forecast results, and supplementary materials are provided in an Online Appendix.

\section{Econometric framework}\label{sec:econometrics}
\subsection{Quantile regression with time-varying parameters}\label{subsec:QR}
Let $\{y_t\}_{t=1}^T$ be a scalar time series and $\{\bm{x}_t\}_{t=1}^T$ a $K\times1$-vector of explanatory variables at time $t=1,\hdots,T$. The explanatory variables may comprise an intercept, observed/latent factors, additional covariates or lags of the endogenous variable. A general version of TVP-QR is given by:
\begin{equation}
y_t = \bm{x}_t'\bm{\beta}_{pt} + \epsilon_t, \quad \text{with} \quad \int_{-\infty}^{0} f_p(\epsilon_t)\text{d}\epsilon_t = p. \label{eq:QR}
\end{equation}
Define $q_p(\bm{x}_t) = \bm{x}_t'\bm{\beta}_{pt}$ as the $p$th quantile regression function of $y_t$ conditional on $\bm{x}_t$, for $p\in(0,1)$. The regression coefficients are collected in a $K\times1$-vector $\{\bm{\beta}_{pt}\}_{t=1}^T$. They vary over time and are specific to the $p$th quantile. The error term $\epsilon_t$ with density $f_p(\bullet)$ has its $p$th quantile equal to zero. Following \citet{yu2001bayesian}, the density $f_p(\bullet)$ is chosen to be the asymmetric Laplace (AL$_p$) distribution. This is due to the correspondence between frequentist and Bayesian inference that this likelihood implies.\footnote{The error term with respect to quantile $p$ is $\epsilon_t\sim\text{AL}_p(\sigma_{pt})$, with density $f_p(\epsilon_t) = p(1-p)/\sigma_{pt}\exp(-\rho_p(\epsilon_t)/\sigma_{pt})$ where $\rho_p(x) = x(p-\mathbb{I}(x<0))$ is the check/loss function and $\mathbb{I}(\bullet)$ is an indicator function which yields one if its argument is true and zero otherwise.} \citet{kozumi2011gibbs} use a mixture representation of the AL$_p$ distribution to cast a constant parameter version of (\ref{eq:QR}) as a conditionally Gaussian model. 

In this paper, in addition to TVPs, I extend the Bayesian QR to feature a time-varying scale parameter similar to a stochastic volatility model.\footnote{Other approaches to introduce conditional heteroskedasticity within quantiles are GARCH-type models \citep[e.g.,][]{gerlach2011bayesian,caporin2018measuring}.} To achieve this, define auxiliary variables $v_{pt}\sim\mathcal{E}(\sigma_{pt})$ which follow an exponential distribution with time-varying scaling $\sigma_{pt}$, and $u_t\sim\mathcal{N}(0,1)$. The model in (\ref{eq:QR}) can be written as:
\begin{equation}
y_t = \bm{x}_t'\bm{\beta}_{pt} + \theta_p v_{pt} + \tau_p\sqrt{\sigma_{pt} v_{pt}}u_t, \quad \theta_p = \frac{1-2p}{p(1-p)}, \quad \tau_p^2 = \frac{2}{p(1-p)}. \label{eq:QR-N}
\end{equation}
In a Gibbs sampling algorithm, the term $\theta_p v_{pt}$ shifts the location of the observations in $y_t$, while $\tau_p\sqrt{\sigma_{pt} v_{pt}}$ reweights them such that $\bm{\beta}_{pt}$ targets the relationship between $y_t$ and $\bm{x}_t$ with respect to the $p$th quantile. To see this more clearly, let $\tilde{y}_{pt} = (y_t - \theta_p v_{pt}) / (\tau_p\sqrt{\sigma_{pt} v_{pt}})$ and $\tilde{\bm{x}}_{pt} = (\tau_p\sqrt{\sigma_{pt} v_{pt}}\bm{I}_K)^{-1}\bm{x}_t$ with $\bm{I}_K$ denoting an identity matrix of size $K$. Conditional on $v_{pt}$ and $\sigma_{pt}$, (\ref{eq:QR-N}) can be written as a standard TVP regression:
\begin{equation}
\tilde{y}_{pt} = \tilde{\bm{x}}_{pt}'\bm{\beta}_{pt} + u_t, \quad u_t\sim\mathcal{N}(0,1).\label{eq:measurement}
\end{equation}
The parameters $\theta_p,\tau_p$ and $v_{pt}$ are the usual parameters shifting and reweighting observations in Bayesian QR. The time-varying scales $\sigma_{pt}$ introduce additional flexibility by allowing for both quantile and time-specific differences in volatilities. This allows the model to discriminate between time-varying signals and noise, and induces a varying degree of smoothness in quantile estimates.

I introduce time-variation in the quantile specific regression coefficients and the logarithmic scale parameters via random walk state equations:
\begin{align}
\bm{\beta}_{pt} &= \bm{\beta}_{pt-1} + \bm{\eta}_{pt}, \quad \bm{\eta}_{pt}\sim\mathcal{N}\left(\bm{0},\bm{\Omega}_{pt}\right)\label{eq:state-eq},\\
\log(\sigma_{pt}) &= \log(\sigma_{pt-1}) + e_{pt}, \quad e_{pt}\sim\mathcal{N}\left(0,\varsigma_p^2\right)\label{eq:scalestate-eq},
\end{align}
with $K\times K$-matrix $\bm{\Omega}_{pt}=\text{diag}\left(\omega_{p1,t},\hdots,\omega_{pK,t}\right)$ collecting independent state innovation variances on its diagonal and $\varsigma_p^2$ refering to the state innovation variance of the scale parameters. Note that combining (\ref{eq:measurement}) and (\ref{eq:state-eq}) yields a state space model with standard normal measurement errors. This enables the use of all standard methods for Gaussian state space models that are available in frameworks specified in terms of the conditional mean.

\subsubsection{Priors for the time-varying coefficients}
Time-variation for the $k$th coefficient in $\bm{\beta}_{pt}$ is governed by $\omega_{pk,t}$ for $k=1,\hdots,K$. There are several options to model these variances, each of them being consequential for the dynamic evolution of the states. The default option is to disregard varying degrees of time variation and rely on a constant specification, that is, $\omega_{pk,1}=\hdots=\omega_{pk,T}=\omega_{pk}$. A common prior in this case is the \textit{inverse Gamma} distribution (subsequently labeled \texttt{iG}). This can be achieved by assuming independent inverse Gamma priors, $\omega_{pk}\sim\mathcal{G}^{-1}(m,n)$. In the empirical application, I choose a weakly informative prior with $m=n=0.1$.

To impose a time-varying degree of shrinkage, the first option is to construct a prior which detects the necessity of time variation on a $t$-by-$t$ basis. The prior I assume is $\omega_{pk,t} = \lambda_{pk}^2\phi_{pk,t}^2$ with $\lambda_{pk}\sim\mathcal{C}^{+}(0,1)$ and $\phi_{pk,t}\sim\mathcal{C}^{+}(0,1)$. $\mathcal{C}^{+}$ refers to the half-Cauchy distribution. This consideration implies that there is no persistent state evolution for $\omega_{pk,t}$, and for this reason I refer to it as the \textit{static horseshoe} prior (labeled \texttt{shs}). In other words, I impose an overall degree of shrinkage $\lambda_{pk}$ towards constancy, with scalings $\phi_{pk,t}$ providing local adaptiveness for periods where shifts in the model parameters are required. This prior is closely related to the one used in \citet{korobilis2021time}.

Persistence in the shrinkage process can be achieved by the \textit{dynamic horseshoe} prior (labeled \texttt{dhs}). Here, I assume that $\omega_{pk,t} = \lambda_{p0}\lambda_{pk}\phi_{pk,t}$ and consider this quantity on the log-scale to obtain a joint law of motion for $\psi_{pk,t} = \log(\lambda_{p0}\lambda_{pk}\phi_{pk,t}) = \log(\omega_{p,kt})$,
\begin{equation}
\psi_{pk,t} = \mu_{\psi,pk} + \varphi_{pk}(\psi_{pk,t-1} - \mu_{\psi,pk}) + \nu_{pk,t}, \quad \nu_{pk,t}\sim\mathcal{Z}(c,d,0,1). \label{eq:shrinkstate}
\end{equation}

Following \citet{kowal2019dynamic}, this respresentation establishes a dynamic version of the horseshoe prior (for $c=d=1/2$), where $\lambda_{p0}$ acts as global shrinkage parameter specific to quantile $p$, $\lambda_{pk}$ is a covariate specific shrinkage parameter, $\phi_{pk,t}$ a covariate and time-specific shrinkage parameter and $\mathcal{Z}$ denotes the Z-distribution. The priors on the shrinkage parameters are $\lambda_{p0}\sim\mathcal{C}^{+}(0,1/TK)$ and $\lambda_{pk}\sim\mathcal{C}^{+}(0,1)$.

\subsubsection{Other priors}
The prior setup is completed by assuming an inverse Gamma prior with $\sigma_p\sim\mathcal{G}^{-1}(a/2,b/2)$ for the case of a time-invariant scale parameter (labeled \texttt{TIS}) of the AL$_p$ distribution (i.e., for $\sigma_{p1}=\hdots=\sigma_{pT} = \sigma_p$), and an inverse gamma prior on the state innovation variance of the logarithmic time-varying process of the scale parameter (labeled \texttt{TVS}), $\varsigma_p^2\sim\mathcal{G}^{-1}(e,f)$. For the empirical application, I set $a = b = 0.1$, $e=3$ and $f=0.3$ to achieve weakly informative priors. 

The resulting posterior distributions and details on the sampling algorithm are provided in Appendix \ref{app:A:Sampling}. Disregarding a number of draws as burn-in, the MCMC algorithm delivers draws from the desired posterior distributions. In the empirical application of this paper, I discard the initial 3,000 draws as burnin and use each third of the 9,000 subsequent draws for posterior and predictive inference.

\subsection{Parallelization and noncrossing quantiles}\label{subsec:noncrossing}
Note that there is no dependence between coefficients across quantiles for QRs as in (\ref{eq:QR}), so estimation for different values of $p$ can easily be parallelized. The recursive forecast exercise shown in this paper, for instance, employs a high-performance cluster which distributes all individual quantile regressions across nodes and collects them afterwards. The \texttt{R}-code associated with this paper also provides routines to distribute the computational burden to any number of available CPUs on a single machine. This feature enables the individual quantile-specific models to be estimated roughly in the same time as a model targeting the unconditional mean. However, this comes with a caveat. Disjoint models of quantiles may be problematic, since independent estimates of the quantile function are not guaranteed to be monotonic.\footnote{The requirement of monotonicity follows from the definition of quantiles. Note that \citet{griffin2020bayesian} or \citet{wu2021bayesian} propose frameworks to model a set of quantiles in a joint econometric framework.}

To ensure that the estimated quantiles are monotonic, i.e., $q_{p_1}(y_t|\bm{x}_t) < q_{p_2}(y_t|\bm{x}_t)$ for $p_1 < p_2$, I rely on an auxiliary Gaussian process regression \citep[GPR, see][]{williams2006gaussian} and post-process the collected MCMC draws for all considered quantiles using the two-step approach discussed in \citet{rodrigues2017regression}. Let $p=\{0.05,0.10,\hdots,0.90,0.95\}$ denote the $P=19$ quantiles of interest, and define $\bm{p}=(0.05,0.10,\hdots,0.90,0.95)'$.\footnote{Note that the approach trivially generalizes to a more granular grid of quantiles, albeit at the cost of additional computational burden.} The approach of \citet{rodrigues2017regression} involves constructing a $P\times P$ matrix
\begin{align*}
\bm{Q}_{t,(p,\tilde p)}(y_t|\bm{x}_t) &= F^{-1}(\tilde p; \bm{x}_t'\hat{\bm{\beta}}_{pt},\hat{\sigma}_{pt}, p) &&\\
&=\begin{cases}
\bm{x}_t'\hat{\bm{\beta}}_{pt} + \frac{\hat{\sigma}_{pt}}{1-p} \log\left(\frac{\tilde p}{p}\right), \quad 
\text{if } 0 \leq \tilde p \leq p,\\
\bm{x}_t'\hat{\bm{\beta}}_{pt} - \frac{\hat{\sigma}_{pt}}{p} \log\left(\frac{1-\tilde p}{1-p}\right), \quad
\text{if } p \leq \tilde p \leq 1.
\end{cases}
\end{align*}
This matrix collects induced quantiles, that is, the quantile function of the fitted AL$_p$ model with respect to the full grid of quantiles $\tilde p=\{0.05,0.10,\hdots,0.90,0.95\}$. The variables $\hat{\bm{\beta}}_{pt}$ and $\hat{\sigma}_{pt}$ refer to the posterior mean estimates of the respective quantities. Consequently, this auxiliary model provides $P-1$ additional estimates of the quantile for each $p$, since the raw quantile estimates $\bm{x}_t'\hat{\bm{\beta}}_{pt}$ are collected on the diagonal of $\bm{Q}_{t,(p,\tilde p)}(y_t|\bm{x}_t)$.

This additional information to unify independent quantile regressions is exploited via the following GPR:
\begin{equation*}
\bm{Q}_{t,(p,\tilde p)}(y_t|\bm{x}_t) = g_t(\bm{p}) + \bm{\varepsilon}_t, \quad g_t(\bm{p})\sim\mathcal{GP}(\bm{0},\bm{\Upsilon}_t), \quad \bm{\varepsilon}_t\sim\mathcal{N}(\bm{0},\bm{\Sigma}_t).
\end{equation*}
The covariance matrix $\bm{\Sigma}_t$ is diagonal and features the posterior variances of the corresponding element $\bm{Q}_{t,(p,\tilde p)}(y_t|\bm{x}_t)$ divided by the number of retained MCMC draws. The Gaussian process covariance is chosen such that it reflects a decreasing function of the distance between quantiles via a squared exponential kernel. The elements of the square matrix $\bm{\Upsilon}_t$, $\upsilon_{t,(p,\tilde p)}$, are thus defined as
\begin{equation*}
\upsilon_{t,(p,\tilde p)} = s^2 \exp\left(-\frac{1}{2w_t^2} (p-\tilde p)^2 \right),
\end{equation*} 
where $w_t$ is the bandwidth and $s^2$ is a variance hyperparameter of the prior. The variance is set to $s^2=100$ to yield a comparatively uninformative prior. The final quantile estimate is given by the fitted values of the GPR at all desired levels of $p$. \citet{rodrigues2017regression} show that this corresponds to a weighted average of the underlying induced quantiles, which is consistent and exhibits favorable empirical properties.

It remains to discuss the choice of the bandwidth $w_t$. \citet{rodrigues2017regression} show that there always exists a bandwidth which guarantees noncrossing quantiles, and note that as $w_t\rightarrow\infty$ one obtains equal weights for the induced quantile, while the case $w_t\rightarrow0$ puts non-zero weights only on the raw quantile estimates. Consequently, I select the minimal $w_t$ that results in noncrossing quantiles for all $p$ and $t=1,\hdots,T$. This approach is subsequently referred to as \texttt{GPt}. An alternative is to choose a $w_t = w$ for all $t$, which is the default setup for constant parameter QRs in \citet{rodrigues2017regression}. The latter choice is labeled \texttt{GP} and typically results in smoother estimates of the quantiles for TVP-QR over time, since $w$ will be the maximum value of the set of minimal values of $\{w_t\}_{t=1}^T$ that guarantee noncrossing quantiles over time.

\section{The unobserved component quantile regression}\label{sec:motivation}
\begin{figure}[t]
  \begin{subfigure}{\textwidth}
  \caption{US}\vspace*{-0.75em}
  \includegraphics[width=\textwidth]{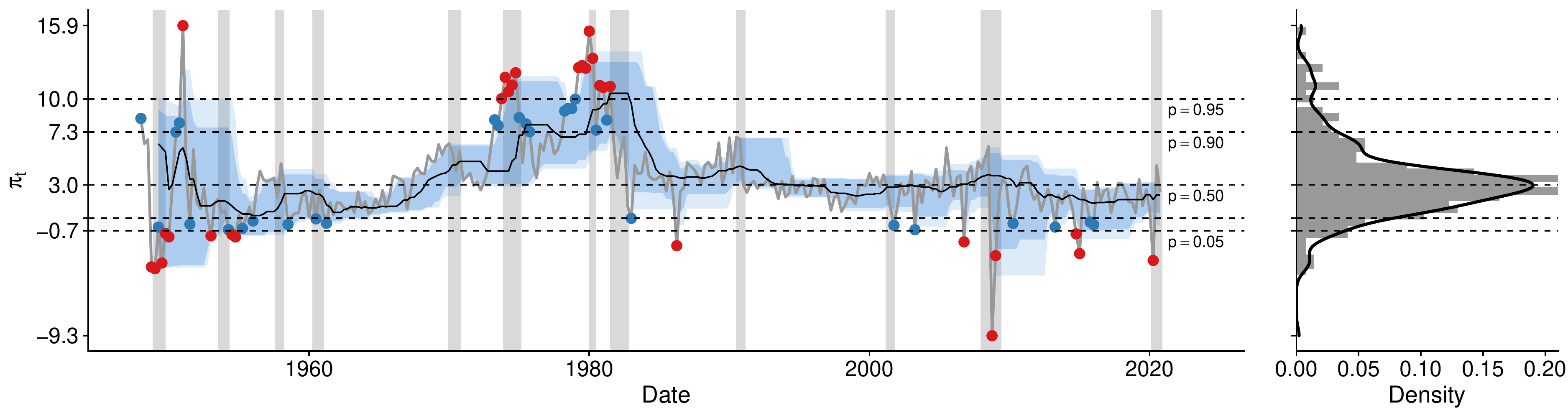}
  \end{subfigure}
  \begin{subfigure}{\textwidth}
  \caption{UK}\vspace*{-0.75em}
  \includegraphics[width=\textwidth]{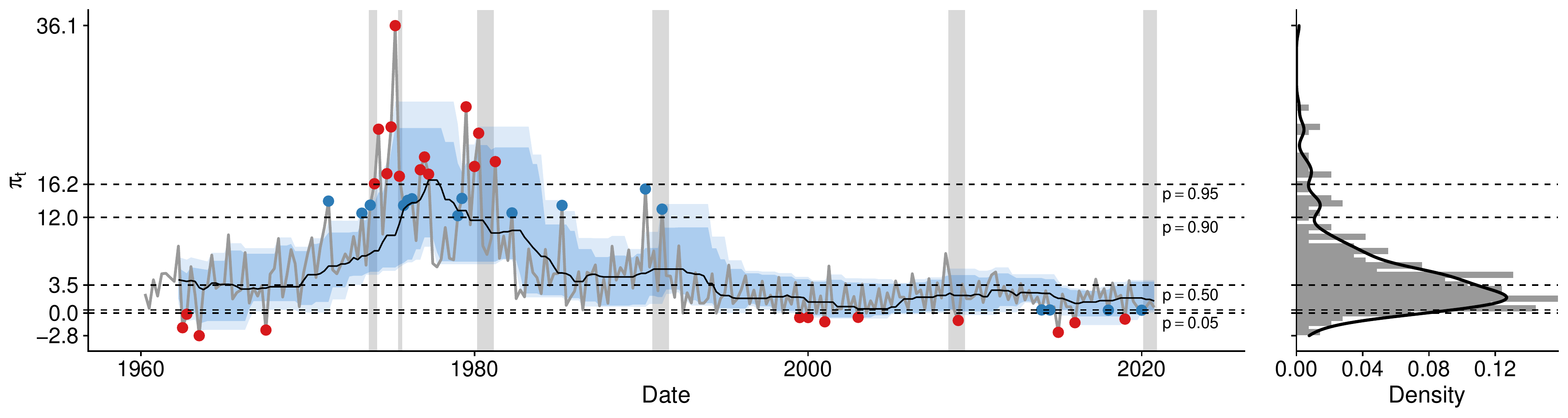}
  \end{subfigure}
  \begin{subfigure}{\textwidth}
  \caption{EA}\vspace*{-0.75em}
  \includegraphics[width=\textwidth]{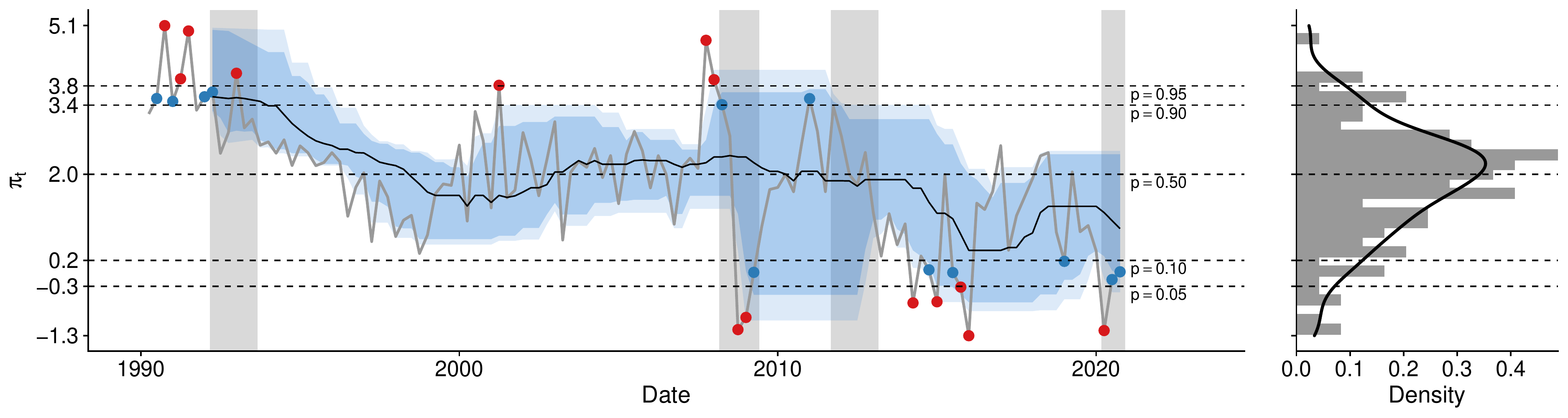}
  \end{subfigure}
  \caption{Inflation series, distribution over time and tail observations.}\label{fig:inflation}\vspace*{-1em}
  \caption*{\scriptsize{\textit{Notes:} Inflation for the United States (US), the United Kingdom (UK) and the euro area (EA) and histogram/smoothed kernel density estimate of observations. The grey shaded areas are recessions. Black dashed horizontal lines mark the quantiles $p\in\{0.05,0.10,0.50,0.90,0.95\}$, blue shaded areas (dark blue: $p=0.10,0.90$; light blue: $p=0.05,0.95$) and solid black line ($p=0.50$) indicate $p$ estimated with a rolling window of $20$ observations (initialized using $8$ observations). Red ($\color{colQRUC}{\bullet}$ with $\pi_t < \pi_{0.05}$ or $\pi_t > \pi_{0.95}$) and blue ($\color{colUCSVM}{\bullet}$ with $\pi_{0.05} \leq \pi_t < \pi_{0.10}$ or $\pi_{0.9} \geq \pi_t < \pi_{0.95}$) dots mark observations in the tails of the distribution, with $\pi_{p}$ denoting the $p$th unconditional quantile over time.}}
\end{figure}

This section first provides descriptive statistics and stylized facts of inflation dynamics, which subsequently motivate the baseline model specification in the empirical part of this paper. For the US and the UK, I use the consumer price index starting 1948:Q1 (US) and 1960:Q1 (UK). Both series are available at \href{https://fred.stlouisfed.org}{fred.stlouisfed.org}. For the EA, I rely on the harmonized index of consumer prices starting 1990:Q1, available at \href{https://sdw.ecb.europa.eu}{sdw.ecb.europa.eu}. All three range until 2020:Q4. Inflation is defined as $\pi_{t} = 400\log(P_t/P_{t-1})$, where $P_{t}$ denotes the price index at time $t$. Figure \ref{fig:inflation} shows inflation for all three economies, alongside a histogram and smoothed density estimate of all observations over time.

Panel (a) of Fig. \ref{fig:inflation} shows $\pi_t$ for the US. The majority of observations above the $90$th and $95$th percentile of the unconditional distribution over time occur during the late $1970$s and early $1980$s. During the $1950$s, a period of high volatility, I observe several values below the $10$th and $5$th percentile. The majority of the remaining observations allocated in the interval $\pi_{0.05} \leq \pi_t < \pi_{0.10}$ occur during or just after recessions until the early $1960$s. After the Volcker chairmanship of the Federal Reserve ended in $1987$, all observations for inflation in the tails of the unconditional distribution are located below the $10$th percentile. This suggests a shift from upside risks of excessive inflation during the $1970$s and $1980$s towards downside risks of deflation in later periods of the sample. This stylized fact may be linked to the decoupling of inflation and inflation volatility discussed in \citet{chan2017stochastic}. Considering the density plot on the right-hand side, the dynamic evolution of inflation translates to a slightly right-skewed unconditional distribution with heavy tails. Investigating the rolling window quantile estimates suggests substantial movements and asymmetries with a varying degree of persistence.

Turning to the UK in panel (b) of Fig. \ref{fig:inflation}, the density plot on the right-hand side shows a substantially right-skewed unconditional distribution. All of these observations in the right tail of the distribution occur between $1975$ and $1990$, a high-volatility period with substantial upside risks of excessive inflation. The peak is reached just before the brief recession in late $1975$. The volatility of inflation in the UK visibly decreased after the European exchange-rate mechanism (ERM) crisis in $1992$. Since then, the only observations in the tails of the distributions are those below the $10$th and $5$th percentile. This shift from upside risk towards downside risk is similar to the case of the US, and may again be linked to policy changes by the Bank of England. Asymmetries in rolling window quantile estimates occur mainly during the late $1970$s to the early $1990$s, while movements in quantiles are muted from the ERM crisis onwards.

A different pattern is visible for the shorter inflation series of the EA, displayed in panel (c) of Fig. \ref{fig:inflation}. Considering the unconditional distribution of inflation over time on the right-hand side, I observe a slightly left-skewed distribution. The median is approximately located at the European Central Bank's target of ``[...] below two percent.'' The maximum value is observed in the early $1990$s, with some values exceeding the $90$th percentile in the buildup to the global financial crisis, and the rebound between the Great Recession and the European sovereign debt crisis. In the years after the latter crisis, several observations fall into the left tail of the distribution. Similarly deflationary quarters are present during the Great Recession, and in early $2020$, at the onset of the recession induced by the Covid-19 pandemic. The rolling window estimate of quantiles suggests only a minor relevance of asymmetries, but indicates that the median of inflation appears to decline. Some periods, such as during the Great Recession and the European sovereign debt crisis feature markedly wider distributions.

The proposed model specification is motivated based on the literature on forecasting the conditional mean of inflation. Popular and successful approaches, particularly for forecasting, are variants of the unobserved component model with stochastic volatility \citep[see, e.g.,][]{stock2007has,chan2013new,chan2017stochastic,chan2018new,jarocinski2018inflation}. The simplest case of this model assumes a persistent (unobserved/latent) trend for the mean, usually augmented with some form of conditional heteroskedasticity. Such models are aimed at providing an accurate model for the conditional mean, which combined with a potentially time-varying variance yields estimates of the conditional quantiles of inflation as a by-product. 

In this paper, I ask how explicit models of quantiles within this class compare to mean-based specifications. Interest centers both on overall forecast performance, but also how the competing models perform for different parts of the distribution. The proposed econometric framework thus extends the conventional unobserved component model for the conditional mean by introducing analogous but explicit unobserved components for different quantiles of the distribution of inflation. 

Figure \ref{fig:inflation} provides a rough gauge of time-variation in the conditional quantiles of inflation. Movements in quantile estimates based on a rolling window clearly indicate the presence of time-variation across all parts of the conditional distribution. And these patterns are sometimes asymmetric, with movements differing in terms of sign, magnitude, timing and persistence. This observation motivates the choice of imposing dynamic shrinkage. Such priors are designed to allow for sudden breaks in the targeted series, they are equipped to rule out time-variation if dynamics are stable, and they also allow for moderate shifts or trending behavior. These features are clearly desirable for the case of inflation, particularly in the US and the UK, with distinct differences in dynamics observable between periods such as the Volcker Disinflation or the Great Moderation.

\subsection{Details on model specification}\label{subsec:UCQR}
\subsubsection{Variants of unobserved component quantile regressions}
I propose to model inflation using the simplest variant of the general TVP-QR model proposed in Section \ref{sec:econometrics}. This model generalizes the UC-SV model of \citet{stock2007has} to modeling the conditional quantiles of inflation. The UCQR model is given by:
\begin{align}
\pi_t &= \alpha_{pt} + \epsilon_t, \quad \epsilon_t\sim\text{AL}_p(\sigma_{pt}),\label{eq:UCQR1}\\
\alpha_{pt} &= \alpha_{pt-1} + \eta_{pt}, \quad \eta_{pt}\sim\mathcal{N}\left(0,\omega_{pt}\right),\label{eq:UCQR3}
\end{align}
with $\epsilon_t$ following an AL$_p$ distribution. I rely on the Gaussian approximation of the AL$_p$ distribution using auxiliary variables discussed in detail in Sub-Section \ref{subsec:QR}. The competing model specifications include both a time-invariant scale (TIS) and a time-varying scale (TVS) version of UCQR.

To assess the robustness of this approach with respect to priors, I consider the dynamic horseshoe (\texttt{dhs}), the static horseshoe (\texttt{shs}) and a conventional inverse Gamma (\texttt{iG}) prior to determine $\omega_{pt}$. The paper also provides a comparison between unprocessed and processed quantile estimates. Unprocessed estimates (i.e., noncrossing of quantiles is not guaranteed) are referred to as \texttt{raw}. They are shown alongside those adjusted using the two-stage GPR procedure discussed in Section \ref{subsec:noncrossing}. \texttt{GP} refers to the bandwidth parameter $w$ being fixed over time, \texttt{GPt} indicates a time-varying $w_t$.

\subsubsection{Mean-based competing models}
The UCQR model is compared to mean-based versions of the UC model. In particular, I consider the following models as competing specifications:
\begin{itemize}[leftmargin=1.5em,itemsep=0em]
  \item[---] \textbf{Unobserved component with stochastic volatility} ({UC-SV})\\
  This model assumes a measurement equation of the form $\pi_t = \alpha_t + \epsilon_t$, with error term $\epsilon_t\sim\mathcal{N}\left(0,\exp(h_t)\right)$, a random walk state equation $\alpha_t = \alpha_{t-1} + \sqrt{\omega_{\alpha t}} e_{\alpha t}$ and log-volatility process $h_t = h_{t-1} + \varsigma_{h} e_{ht}$. The errors of the state equations follow independent standard normal distributions, $e_{st}\sim\mathcal{N}(0,1)$ for $s\in\{\alpha,h\}$. This is a variant of the model proposed by \citet{stock2007has}, and the natural competitor to UCQR given that it is specified analogously, but targets solely the conditional mean rather than conditional quantiles.
  \item[---] \textbf{Unobserved component with stochastic volatility in mean} ({UC-SVM})\\ 
  This model assumes a measurement equation of the form $\pi_t = \alpha_t + \gamma_t \exp(h_t) + \epsilon_t$ with error term $\epsilon_t\sim\mathcal{N}\left(0,\exp(h_t)\right)$. The state equations are $\alpha_t = \alpha_{t-1} + \sqrt{\omega_{\alpha t}} e_{\alpha t}$, $\gamma_t = \gamma_{t-1} + \sqrt{\omega_{\gamma t}} e_{\gamma t}$, and for the stochastic volatilities $h_t = h_{t-1} + \varsigma_{h} e_{ht}$, with $e_{st}$ following independent standard normal distributions for $s\in\{\alpha,\gamma,h\}$. This specification was originally used in \citet{chan2017stochastic}, and establishes a time-varying correlation between the state and measurement equation. It thus allows for flexible feedback effects between inflation and its volatility, relating to the literature on financial conditions and uncertainty. A multivariate version of this model in the context of tail risks in GDP growth has been used by \citet{caldara2020macroeconomic}.
\end{itemize}
UC-SV and UC-SVM are implemented and estimated as described in \citet{huber2020dynamic} with a dynamic horseshoe prior. An overview of all model specifications and adjustments is provided in Table \ref{tab:models}. The forecast exercise thus features three competing models alongside $18$ variants of UCQR governed by the respective prior, treatment of the scale parameter and ex post adjustment of quantile estimates.

\begin{table*}[!t]
\caption{Model overview.}\vspace*{-1.5em}
\begin{center}
\begin{small}
\begin{threeparttable}
\begin{tabular*}{\textwidth}{@{\extracolsep{\fill}} llll}
\toprule
\textbf{Model} & \textbf{Prior} & \textbf{Volatility/Scale} & \textbf{Adjustment}\\
\midrule
\texttt{UC} & \texttt{dhs} & \texttt{SV}, \texttt{SVM} & --- \\
\texttt{UCQR} & \texttt{dhs}, \texttt{shs}, \texttt{iG} & \texttt{TIS}, \texttt{TVS} & \texttt{raw}, \texttt{GP}, \texttt{GPt} \\
\bottomrule
\end{tabular*}
\begin{tablenotes}[para,flushleft]
\scriptsize{\textit{Notes}: Unobserved component (\texttt{UC}), unobserved component quantile regression (\texttt{UCQR}); dynamic horseshoe (\texttt{dhs}), static horseshoe (\texttt{shs}), inverse Gamma (\texttt{iG}) prior; stochastic volatility (\texttt{SV}), stochastic volatility in mean (\texttt{SVM}), time-invariant scale (\texttt{TIS}), time-varying scale (\texttt{TVS}) parameter; \texttt{raw} refers to unprocessed estimates of the conditional quantiles, \texttt{GP} is adjustment using the Gaussian process regression with the bandwidth $w$ fixed over time, \texttt{GPt} marks the Gaussian process regression with time-varying bandwidth $w_t$.}
\end{tablenotes}
\end{threeparttable}
\end{small}
\end{center}
\label{tab:models}
\end{table*}

\section{Conditional mean and quantile based models for inflation}\label{sec:insample}
\begin{figure}[!t]
  \begin{subfigure}{\textwidth}
  \caption{US}\vspace*{-0.75em}
  \includegraphics[width=\textwidth]{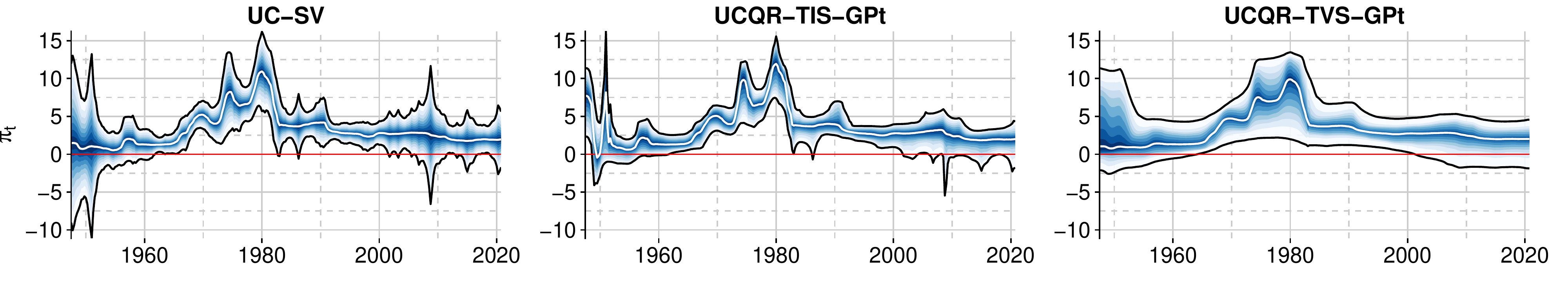}
  \end{subfigure}
  \begin{subfigure}{\textwidth}
  \caption{UK}\vspace*{-0.75em}
  \includegraphics[width=\textwidth]{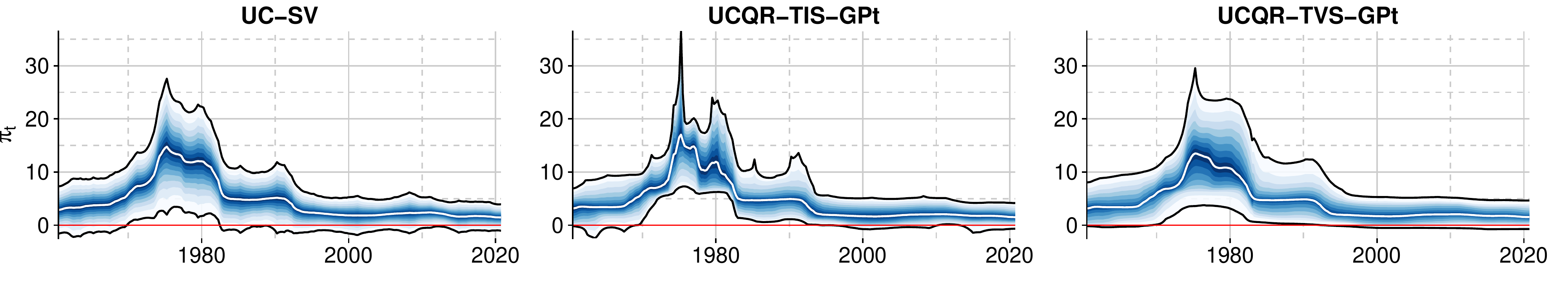}
  \end{subfigure}
  \begin{subfigure}{\textwidth}
  \caption{EA}\vspace*{-0.75em}
  \includegraphics[width=\textwidth]{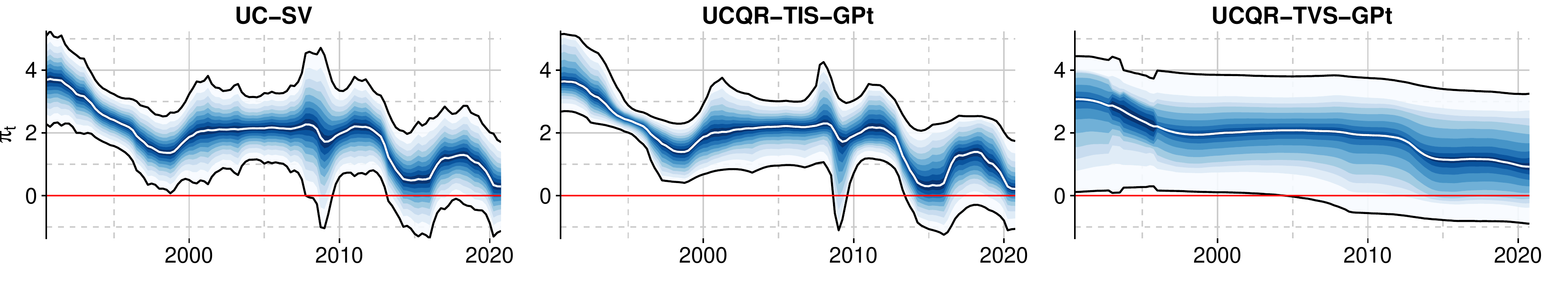}
  \end{subfigure}
  \caption{Unobserved components for UC-SV and variants of UCQR with a dynamic horseshoe.}\label{fig:insample}\vspace*{-1em}
  \caption*{\scriptsize{\textit{Notes:} Unobserved component model with stochastic volatility (UC-SV) and unobserved component quantile regression (UCQR) with time-invariant (TIS) or time-varying (TVS) scale parameter, adjusted using a Gaussian process regression with time-varying bandwidth (GPt). The red line marks zero. The solid black lines are the posterior mean of the $5$th and $95$th quantiles, the solid white line is the $50$th percentile (median). The blue shaded areas cover quantile pairs (e.g., $10$th to $90$th percentile) in increments of five.}}
\end{figure}

Figure \ref{fig:insample} compares in-sample findings for the US, the UK and the EA. The red line marks zero, the solid black lines are the posterior mean of the $5$th and $95$th quantiles, and the solid white line is the $50$th percentile (median). The blue shaded areas cover quantile pairs (e.g., $10$th to $90$th percentile) in increments of five. Variants of UCQR with TIS and TVS, post-processed using a GPR with time-varying bandwidth are shown alongside the conventional UC-SV model. Additional empirical results are provided in the Online Appendix. I consider the posterior mean of the quantile estimates for the UCQR-variants, and rely on MCMC output based on $\pi_t\sim\mathcal{N}(\alpha_t,\exp(h_t))$ for UC-SV.\footnote{In a recent paper, \citet{wu2021bayesian} note that approaches using the AL likelihood yield invalid asymptotic variances of posterior distributions, and thus produce incorrect inference from a frequentist perspective. Their findings are irrelevant for the purposes of this paper, since both the discussion of in-sample findings and the forecast evaluation metrics require only a point estimate from the posterior distribution of quantiles, and I am not concerned with more intricate inferential problems such as hypothesis testing.}

The charts in Fig. \ref{fig:insample} show several interesting differences with respect to the three models and over the cross-section of economies. Most strikingly for the case of the US, UC-SV results in approximately symmetric estimates of upside and downside risk, apparent mainly in the early part of the sample and during the Great Recession. While the mean-based model suggests that both risks are about equally likely, the implied distributions from quantile-based models early in the sample are skewed, and point mainly towards risks of inflation rather than deflation. This situation is mirrored during the Great Recession, with a comparatively stable upper tail of the distribution and elevated risk of deflation for UCQR-TIS-GPt, and a symmetric widening of the distribution in the case of UC-SV. The version of UCQR with TVS overall results in much smoother and persistent estimates of quantiles, particularly for the lower tails during the late $1970$s and early $1980$s.

Estimates for the quantiles of inflation in the UK provide a similar picture compared to the US, albeit with the difference of UC-SV and UCQR-TVS-GPt exhibiting a rather similar dynamic evolution. By contrast, estimates for the case of UCQR-TIS-GPt are much noisier, particularly during the high-inflation period during the late $1970$s. The UK provides a prime example to investigate the empirical features of the dynamic horseshoe prior, taking the median as an example. First, a modest degree of time-variation is visible, with a slight upward trend of inflation peaking around $1975$. Second, there are several abrupt changes, particularly around $1980$ and $1992$, with substantial decreases of inflation. These large breaks look similar to what one would obtain from running a mixture-innovation or threshold model. Third, there are prolonged periods of virtually no time-variation, e.g., between $1981$ and $1990$, or after $1995$.

For the case of the EA, different to the US and UK, UC-SV and UCQR-TIS-GPt result in rather similar estimates of the quantiles of inflation over time. UCQR-TVS-GPt, however, looks very different to the former two models, with much smoother estimates for all considered quantiles. In the early part of the sample, the implied distribution is much wider and substantially skewed, and movements particularly in the tails are muted. In fact, the lower tail does not exhibit time-variation at all until around 2005. This again showcases the effects of imposing dynamic shrinkage. 

Summarizing, two key findings are worth noting. First, mean-based models typically indicate symmetric risks, while UCQR allows for asymmetric movements in quantiles and thus skewed and heavily tailed distributions. This finding arises from defining distinct quantile regressions with individual loss functions that specifically target different parts of the distribution, rather than average considerations solely based on the conditional mean. Second, using a time-varying scale parameter for the AL distribution usually results in much smoother estimates of the conditional quantiles. Econometrically, this can be explained by noting that larger values of $\sigma_{pt}$ downweight extreme observations. Relaxing the assumption of conditionally homoskedastic errors in conjunction with a flexible model for the conditional quantiles allows the model to decide whether it fits the underlying series aggressively or puts such dynamics in the stochastic error component $\epsilon_t$. Conditional heteroskedasticity clearly captures part of the movements in the underlying quantiles, thereby leaving less information to fit via the time-varying intercept by quantile.

\section{Predictive inference}\label{sec:forecasts}
To assess predictive accuracy of the competing models, I rely on a pseudo out-of-sample exercise using an expanding window. The sample is split into training and holdout periods. I estimate the models using data from the first training sample to produce $h$-steps ahead forecasts, and evaluate these using the corresponding realized values in the holdout. Subsequently, an additional observation from the holdout is added to the training sample and the procedure is iterated until the holdout is exhausted. The initial sample is chosen such that it consists of $50$ quarterly observations for all three economies, resulting in differently sized holdout periods.

\subsection{Forecast metrics and results}
I consider the quantile weighted cumulative ranked probability score (CRPS) to measure the overall accuracy of the density forecasts at different points of the distribution, alongside conventional log predictive scores (LPSs). Let $\pi_{t+h}^{(fp)}$ denote the point forecast of quantile $p$ at time $t$ for horizon $h$, and $\pi_{t+h}^{(r)}$ is the realization of the inflation measure in the holdout. For {UC-SV} and {UC-SVM}, $\pi_{t+h}^{(fp)}$ is the $p$th quantile of the predictive distribution. For {UCQR}, I obtain forecasts for the $p$th quantile in three variants: the unprocessed mean of the predictive distribution, the output of the GPR with the bandwith $w$ fixed for all $t$ ({UCQR-GP}), and a GPR with $w_t$ selected $t$-by-$t$ ({UCQR-GPt}).

I first define the quantile score \citep[QS, see, e.g.,][]{giacomini2005evaluation,gneiting2007strictly} for quantile $p$ at time $t$ for $h$-steps ahead, which is computed as:
\begin{equation*}
\text{QS}_{p,t+h} = \left(\pi_{t+h}^{(r)} - \pi_{t+h}^{(fp)}\right) \times \left(p - \mathbb{I}\left(\pi_{t+h}^{(r)}<\pi_{t+h}^{(fp)}\right)\right),
\end{equation*}
where $\mathbb{I}$ is the usual indicator function. Raw values for QSs to assess tail forecast performance at specific quantiles are provided in Appendix \ref{app:B:Results}. Quantile weighted CRPSs, following \citet{gneiting2011comparing}, are defined as:
\begin{equation*}
\text{CRPS}_{t+h}(\mathfrak{w}_p) = \int_{0}^{1} \mathfrak{w}_{p} \text{QS}_{p,t+h} \text{d}p,
\end{equation*}
with non-negative weights $\mathfrak{w}_{p}$ on the unit interval putting emphasis on specific parts of the distribution. Four weighting schemes are considered for CRPSs: (a) equal $\mathfrak{w}_{p} = 1$, (b) tails $\mathfrak{w}_{p} = (2p-1)^2$, (c) left tail $\mathfrak{w}_{p} = (1-p)^2$, (d) right tail $\mathfrak{w}_{p} = p^2$, for $p\in\{0.05,0.10,\hdots,0.90,0.95\}$.

To recover a smooth estimate of the entire predictive distribution from the individual QRs, I apply kernel smoothing based on a Gaussian kernel on the estimated quantiles \citep[see also][]{gaglianone2012constructing,korobilis2017quantile}. This procedure can be exploited to compute log predictive scores for the QR-based models.

\begin{table*}[t]
\begin{center}
\begin{scriptsize}
\begin{threeparttable}
\begin{tabular*}{\textwidth}{@{\extracolsep{\fill}} lrrrrrrrrrrrrrrr}
  \toprule
 & \multicolumn{12}{c}{\textbf{CRPS}} &  \multicolumn{3}{c}{\textbf{LPS}} \\
 \cmidrule(l){2-13}
 & \multicolumn{3}{c}{None} & \multicolumn{3}{c}{Tails} & \multicolumn{3}{c}{Left} & \multicolumn{3}{c}{Right} & \multicolumn{3}{c}{} \\
 \cmidrule(l){2-4}\cmidrule(l){5-7}\cmidrule(l){8-10}\cmidrule(l){11-13}\cmidrule(l){14-16}
\multicolumn{1}{r}{$h=$} & 1 & 4 & 12 & 1 & 4 & 12 & 1 & 4 & 12 & 1 & 4 & 12 & 1 & 4 & 12 \\
  \midrule
UC-SV & \colBench$0.48$ & \colBench$0.64$ & \colBench$0.81$ & \colBench$0.10$ & \colBench$0.14$ & \colBench$0.18$ & \colBench$0.15$ & \colBench$0.19$ & \colBench$0.24$ & \colBench$0.14$ & \colBench$0.19$ & \colBench$0.25$ & \colBench$-1.85$ & \colBench$-2.03$ & \colBench$-2.17$ \\ 
  UC-SVM & \colBest$0.96$ & $0.98$ & $1.06$ & \colBest$0.95$ & $1.00$ & $1.09$ & \colBest$0.97$ & $1.01$ & $1.08$ & \colBest$0.94$ & \colBest$0.96$ & $1.05$ & $ 0.03$ & $ 0.07$ & $ 0.09$ \\ 
  \midrule
  \multicolumn{16}{c}{\textbf{UCQR-TIS}} \\
  \midrule
  dhs & $1.02$ & $1.01$ & $0.99$ & $1.02$ & $1.02$ & $1.01$ & $1.02$ & $1.00$ & $0.98$ & $1.02$ & $1.02$ & $1.00$ & $ 0.07$ & $ 0.06$ & $ 0.10$ \\ 
  shs & $1.00$ & $1.01$ & $1.01$ & $1.01$ & $1.03$ & $1.04$ & $1.01$ & $1.03$ & $1.03$ & $0.99$ & $1.00$ & $1.01$ & $ 0.05$ & $ 0.01$ & $-0.04$ \\ 
  iG & $1.01$ & $1.04$ & $1.08$ & $1.04$ & $1.11$ & $1.15$ & $1.02$ & $1.08$ & $1.11$ & $1.01$ & $1.03$ & $1.07$ & $ 0.09$ & $ 0.14$ & \colBest$ 0.18$ \\ 
  dhs-GP & $1.02$ & $1.01$ & $0.99$ & $1.02$ & $1.02$ & $1.01$ & $1.02$ & $1.00$ & $0.98$ & $1.02$ & $1.02$ & $1.00$ & $ 0.07$ & $ 0.08$ & $ 0.12$ \\ 
  shs-GP & $1.00$ & $1.01$ & $1.02$ & $1.00$ & $1.03$ & $1.04$ & $1.01$ & $1.03$ & $1.03$ & $0.99$ & $1.00$ & $1.01$ & $ 0.06$ & $-0.04$ & $ 0.03$ \\ 
  iG-GP & $1.00$ & $1.01$ & $1.02$ & $1.00$ & $1.03$ & $1.04$ & $1.01$ & $1.03$ & $1.03$ & $0.99$ & $1.00$ & $1.01$ & $ 0.04$ & $-0.01$ & $-0.01$ \\ 
  dhs-GPt & $1.01$ & $1.04$ & $1.08$ & $1.05$ & $1.12$ & $1.14$ & $1.02$ & $1.09$ & $1.12$ & $1.01$ & $1.02$ & $1.07$ & $ 0.06$ & \colBest$ 0.16$ & $ 0.09$ \\ 
  shs-GPt & $1.01$ & $1.04$ & $1.08$ & $1.04$ & $1.10$ & $1.14$ & $1.02$ & $1.08$ & $1.11$ & $1.01$ & $1.02$ & $1.06$ & \colBest$ 0.11$ & $ 0.15$ & $ 0.08$ \\ 
  iG-GPt & $1.02$ & $1.01$ & $0.99$ & $1.02$ & $1.02$ & $1.01$ & $1.02$ & $1.00$ & $0.98$ & $1.02$ & $1.02$ & $1.00$ & $ 0.04$ & $ 0.07$ & $ 0.13$ \\ 
  \midrule
  \multicolumn{16}{c}{\textbf{UCQR-TVS}} \\
  \midrule
  dhs & $1.14$ & $1.01$ & $0.93$ & $1.13$ & $1.00$ & $0.90$ & $1.09$ & $0.97$ & $0.91$ & $1.19$ & $1.05$ & $0.94$ & $-0.15$ & $-0.08$ & $-0.21$ \\ 
  shs & $1.13$ & $1.02$ & $0.95$ & $1.10$ & $0.99$ & $0.91$ & $1.07$ & $0.98$ & $0.92$ & $1.19$ & $1.05$ & $0.96$ & $-0.10$ & $-0.10$ & $-0.11$ \\ 
  iG & $1.09$ & $0.98$ & $0.93$ & $1.07$ & $0.95$ & $0.88$ & $1.06$ & \colBest$0.95$ & $0.91$ & $1.12$ & $1.00$ & $0.93$ & $-0.09$ & $-0.10$ & $-0.09$ \\ 
  dhs-GP & $1.14$ & $1.01$ & $0.93$ & $1.13$ & $1.00$ & $0.90$ & $1.09$ & $0.97$ & \colBest$0.91$ & $1.20$ & $1.05$ & $0.94$ & $-0.16$ & $-0.09$ & $-0.19$ \\ 
  shs-GP & $1.12$ & $1.00$ & $0.94$ & $1.10$ & $0.98$ & $0.90$ & $1.07$ & $0.97$ & $0.91$ & $1.16$ & $1.03$ & $0.95$ & $-0.12$ & $-0.13$ & $-0.16$ \\ 
  iG-GP & $1.12$ & $1.00$ & $0.94$ & $1.10$ & $0.98$ & $0.90$ & $1.07$ & $0.97$ & $0.91$ & $1.16$ & $1.03$ & $0.94$ & $-0.13$ & $-0.12$ & $-0.17$ \\ 
  dhs-GPt & $1.09$ & \colBest$0.98$ & \colBest$0.93$ & $1.07$ & $0.95$ & $0.88$ & $1.06$ & $0.95$ & $0.91$ & $1.12$ & $1.00$ & $0.93$ & $-0.09$ & $-0.08$ & $-0.13$ \\ 
  shs-GPt & $1.10$ & $0.98$ & $0.93$ & $1.08$ & \colBest$0.95$ & \colBest$0.88$ & $1.07$ & $0.96$ & $0.92$ & $1.13$ & $1.00$ & \colBest$0.93$ & $-0.13$ & $-0.11$ & $-0.10$ \\ 
  iG-GPt & $1.17$ & $1.03$ & $0.95$ & $1.15$ & $1.03$ & $0.93$ & $1.11$ & $0.99$ & $0.93$ & $1.23$ & $1.08$ & $0.97$ & $-0.16$ & $-0.13$ & $-0.16$ \\ 
   \bottomrule
\end{tabular*}
\begin{tablenotes}[para,flushleft]
\scriptsize{\textit{Notes}: For model abbreviations, see Section \ref{subsec:UCQR} and Tab. \ref{tab:models}. Quantile weighted continuous ranked probability scores (CRPS) for different weighting schemes. The benchmark shows actual values shaded in light grey, all other models are shown relative to the benchmark. The best performing model is indicated in dark grey shading.}
\end{tablenotes}
\end{threeparttable}
\end{scriptsize}
\end{center}
\caption{Density forecast metrics for US inflation.} 
\label{tab:US-dens}
\end{table*}

\begin{table*}[t]
\begin{center}
\begin{scriptsize}
\begin{threeparttable}
\begin{tabular*}{\textwidth}{@{\extracolsep{\fill}} lrrrrrrrrrrrrrrr}
  \toprule
 & \multicolumn{12}{c}{\textbf{CRPS}} &  \multicolumn{3}{c}{\textbf{LPS}} \\
 \cmidrule(l){2-13}
 & \multicolumn{3}{c}{None} & \multicolumn{3}{c}{Tails} & \multicolumn{3}{c}{Left} & \multicolumn{3}{c}{Right} & \multicolumn{3}{c}{} \\
 \cmidrule(l){2-4}\cmidrule(l){5-7}\cmidrule(l){8-10}\cmidrule(l){11-13}\cmidrule(l){14-16}
\multicolumn{1}{r}{$h=$} & 1 & 4 & 12 & 1 & 4 & 12 & 1 & 4 & 12 & 1 & 4 & 12 & 1 & 4 & 12 \\
  \midrule
UC-SV & \colBench$0.93$ & \colBench$0.97$ & \colBench$1.02$ & \colBench$0.19$ & \colBench$0.20$ & \colBench$0.21$ & \colBench$0.26$ & \colBench$0.27$ & \colBench$0.29$ & \colBench$0.30$ & \colBench$0.32$ & \colBench$0.32$ & \colBench$-2.49$ & \colBench$-2.49$ & \colBench$-2.55$ \\ 
  UC-SVM & $1.01$ & $1.00$ & $1.03$ & \colBest$0.98$ & $0.98$ & $1.05$ & $1.01$ & $1.00$ & $1.07$ & \colBest$0.99$ & $0.99$ & $1.00$ & $ 0.04$ & $ 0.05$ & $ 0.02$ \\ 
  \midrule
  \multicolumn{16}{c}{\textbf{UCQR-TIS}} \\
  \midrule
  dhs & $1.00$ & $1.03$ & $1.02$ & $1.00$ & $1.03$ & $1.04$ & $0.99$ & $1.02$ & $1.01$ & $1.01$ & $1.04$ & $1.03$ & \colBest$ 0.17$ & $ 0.16$ & $ 0.05$ \\ 
  shs & \colBest$0.99$ & $1.00$ & $1.00$ & $1.00$ & $1.01$ & $1.04$ & \colBest$0.98$ & $0.99$ & $1.01$ & $1.00$ & $1.01$ & $1.00$ & $ 0.15$ & $ 0.16$ & $ 0.08$ \\ 
  iG & $1.03$ & \colBest$0.96$ & $0.99$ & $1.07$ & $0.98$ & $1.05$ & $1.01$ & \colBest$0.97$ & $1.04$ & $1.06$ & $0.96$ & $0.97$ & $ 0.10$ & $ 0.19$ & $ 0.09$ \\ 
  dhs-GP & $1.00$ & $1.03$ & $1.02$ & $1.00$ & $1.03$ & $1.04$ & $0.99$ & $1.02$ & $1.01$ & $1.01$ & $1.04$ & $1.03$ & $ 0.17$ & $ 0.17$ & $ 0.05$ \\ 
  shs-GP & $0.99$ & $1.00$ & $1.00$ & $1.00$ & $1.01$ & $1.04$ & $0.98$ & $0.99$ & $1.01$ & $1.00$ & $1.01$ & $1.00$ & $ 0.16$ & $ 0.16$ & $ 0.08$ \\ 
  iG-GP & $0.99$ & $1.00$ & $1.00$ & $1.00$ & $1.01$ & $1.04$ & $0.98$ & $0.99$ & $1.01$ & $1.00$ & $1.01$ & $1.00$ & $ 0.13$ & $ 0.16$ & $ 0.05$ \\ 
  dhs-GPt & $1.04$ & $0.96$ & $0.99$ & $1.08$ & \colBest$0.97$ & $1.05$ & $1.01$ & $0.97$ & $1.04$ & $1.07$ & \colBest$0.96$ & $0.97$ & $ 0.03$ & \colBest$ 0.22$ & \colBest$ 0.12$ \\ 
  shs-GPt & $1.04$ & $0.97$ & $1.00$ & $1.06$ & $0.98$ & $1.04$ & $1.02$ & $0.97$ & $1.05$ & $1.06$ & $0.97$ & \colBest$0.97$ & $-0.02$ & $ 0.21$ & $ 0.11$ \\ 
  iG-GPt & $1.00$ & $1.03$ & $1.02$ & $1.00$ & $1.03$ & $1.04$ & $0.99$ & $1.02$ & $1.01$ & $1.01$ & $1.04$ & $1.03$ & $ 0.17$ & $ 0.17$ & $ 0.08$ \\ 
  \midrule
  \multicolumn{16}{c}{\textbf{UCQR-TVS}} \\
  \midrule
  dhs & $1.02$ & $1.05$ & $1.00$ & $1.04$ & $1.07$ & $1.03$ & $1.01$ & $1.03$ & $0.96$ & $1.04$ & $1.08$ & $1.04$ & $ 0.05$ & $ 0.10$ & $ 0.02$ \\ 
  shs & $1.02$ & $1.02$ & $0.98$ & $1.02$ & $1.02$ & $0.99$ & $1.01$ & $1.02$ & $0.97$ & $1.02$ & $1.03$ & $0.99$ & $ 0.07$ & $ 0.09$ & $ 0.05$ \\ 
  iG & $1.01$ & $1.00$ & $0.96$ & $1.01$ & $1.01$ & $0.98$ & $1.00$ & $0.99$ & $0.95$ & $1.02$ & $1.02$ & $0.98$ & $ 0.09$ & $ 0.10$ & $ 0.07$ \\ 
  dhs-GP & $1.02$ & $1.05$ & $1.00$ & $1.04$ & $1.07$ & $1.03$ & $1.01$ & $1.03$ & $0.96$ & $1.04$ & $1.08$ & $1.04$ & $ 0.06$ & $ 0.07$ & $ 0.02$ \\ 
  shs-GP & $1.01$ & $1.03$ & $0.98$ & $1.02$ & $1.03$ & $1.00$ & $1.00$ & $1.01$ & $0.95$ & $1.02$ & $1.04$ & $1.00$ & $ 0.11$ & $ 0.11$ & $ 0.04$ \\ 
  iG-GP & $1.01$ & $1.03$ & $0.98$ & $1.02$ & $1.03$ & $1.00$ & $1.00$ & $1.01$ & $0.96$ & $1.02$ & $1.04$ & $1.00$ & $ 0.10$ & $ 0.13$ & $ 0.03$ \\ 
  dhs-GPt & $1.01$ & $1.00$ & \colBest$0.96$ & $1.01$ & $1.01$ & $0.99$ & $1.00$ & $0.99$ & \colBest$0.95$ & $1.01$ & $1.02$ & $0.98$ & $ 0.12$ & $ 0.13$ & $ 0.07$ \\ 
  shs-GPt & $1.01$ & $1.00$ & $0.97$ & $1.00$ & $1.00$ & \colBest$0.98$ & $1.00$ & $1.00$ & $0.96$ & $1.01$ & $1.00$ & $0.98$ & $ 0.07$ & $ 0.12$ & $ 0.05$ \\ 
  iG-GPt & $1.02$ & $1.06$ & $1.00$ & $1.05$ & $1.08$ & $1.04$ & $1.01$ & $1.03$ & $0.96$ & $1.04$ & $1.09$ & $1.05$ & $ 0.09$ & $ 0.05$ & $ 0.02$ \\ 
   \bottomrule
\end{tabular*}
\begin{tablenotes}[para,flushleft]
\scriptsize{\textit{Notes}: For model abbreviations, see Section \ref{subsec:UCQR} and Tab. \ref{tab:models}. Quantile weighted continuous ranked probability scores (CRPS) for different weighting schemes. The benchmark shows actual values shaded in light grey, all other models are shown relative to the benchmark. The best performing model is indicated in dark grey shading.}
\end{tablenotes}
\end{threeparttable}
\end{scriptsize}
\end{center}
\caption{Density forecast metrics for UK inflation.} 
\label{tab:UK-dens}
\end{table*}

\begin{table*}[t]
\begin{center}
\begin{scriptsize}
\begin{threeparttable}
\begin{tabular*}{\textwidth}{@{\extracolsep{\fill}} lrrrrrrrrrrrrrrr}
  \toprule
 & \multicolumn{12}{c}{\textbf{CRPS}} &  \multicolumn{3}{c}{\textbf{LPS}} \\
 \cmidrule(l){2-13}
 & \multicolumn{3}{c}{None} & \multicolumn{3}{c}{Tails} & \multicolumn{3}{c}{Left} & \multicolumn{3}{c}{Right} & \multicolumn{3}{c}{} \\
 \cmidrule(l){2-4}\cmidrule(l){5-7}\cmidrule(l){8-10}\cmidrule(l){11-13}\cmidrule(l){14-16}
\multicolumn{1}{r}{$h=$} & 1 & 4 & 12 & 1 & 4 & 12 & 1 & 4 & 12 & 1 & 4 & 12 & 1 & 4 & 12 \\
  \midrule
UC-SV & \colBench$0.30$ & \colBench$0.37$ & \colBench$0.43$ & \colBench$0.06$ & \colBench$0.08$ & \colBench$0.10$ & \colBench$0.09$ & \colBench$0.12$ & \colBench$0.15$ & \colBench$0.09$ & \colBench$0.11$ & \colBench$0.12$ & \colBench$-1.42$ & \colBench$-1.91$ & \colBench$-1.59$ \\ 
  UC-SVM & \colBest$0.97$ & $1.06$ & $1.05$ & \colBest$0.96$ & $1.06$ & $1.03$ & \colBest$0.99$ & $1.07$ & $1.02$ & \colBest$0.95$ & $1.05$ & $1.08$ & \colBest$ 0.06$ & $ 0.30$ & $-0.02$ \\ 
  \midrule
  \multicolumn{16}{c}{\textbf{UCQR-TIS}} \\
  \midrule
  dhs & $1.04$ & $1.03$ & $1.00$ & $1.05$ & $1.06$ & $1.01$ & $1.05$ & $1.05$ & $1.01$ & $1.02$ & $1.01$ & $0.98$ & $-0.03$ & $ 0.12$ & \colBest$ 0.05$ \\ 
  shs & $1.02$ & $1.05$ & $1.03$ & $1.03$ & $1.09$ & $1.04$ & $1.05$ & $1.07$ & $1.04$ & $1.00$ & $1.04$ & $1.02$ & $ 0.03$ & $ 0.31$ & $ 0.04$ \\ 
  iG & $1.01$ & $1.06$ & $1.08$ & $1.04$ & $1.13$ & $1.13$ & $1.05$ & $1.08$ & $1.11$ & $0.98$ & $1.05$ & $1.06$ & $ 0.01$ & $ 0.39$ & $-0.20$ \\ 
  dhs-GP & $1.04$ & $1.03$ & $1.00$ & $1.05$ & $1.06$ & $1.01$ & $1.06$ & $1.05$ & $1.01$ & $1.03$ & $1.01$ & $0.98$ & $-0.07$ & $ 0.12$ & $ 0.03$ \\ 
  shs-GP & $1.02$ & $1.05$ & $1.03$ & $1.03$ & $1.09$ & $1.04$ & $1.05$ & $1.07$ & $1.04$ & $1.00$ & $1.04$ & $1.02$ & $ 0.03$ & $ 0.30$ & $ 0.03$ \\ 
  iG-GP & $1.02$ & $1.05$ & $1.03$ & $1.03$ & $1.09$ & $1.04$ & $1.05$ & $1.07$ & $1.04$ & $1.00$ & $1.04$ & $1.02$ & $ 0.03$ & $ 0.30$ & $ 0.04$ \\ 
  dhs-GPt & $1.01$ & $1.06$ & $1.08$ & $1.04$ & $1.13$ & $1.14$ & $1.05$ & $1.09$ & $1.11$ & $0.98$ & $1.06$ & $1.06$ & $ 0.01$ & \colBest$ 0.40$ & $-0.14$ \\ 
  shs-GPt & $1.01$ & $1.06$ & $1.08$ & $1.05$ & $1.13$ & $1.13$ & $1.05$ & $1.09$ & $1.11$ & $0.98$ & $1.05$ & $1.06$ & $-0.07$ & $ 0.29$ & $-0.01$ \\ 
  iG-GPt & $1.04$ & $1.03$ & $0.99$ & $1.05$ & $1.06$ & $1.01$ & $1.06$ & $1.05$ & $1.01$ & $1.02$ & $1.01$ & $0.98$ & $-0.06$ & $ 0.12$ & $ 0.03$ \\ 
  \midrule
  \multicolumn{16}{c}{\textbf{UCQR-TVS}} \\
  \midrule
  dhs & $1.11$ & \colBest$0.95$ & $0.88$ & $1.10$ & $0.95$ & $0.84$ & $1.13$ & $0.96$ & $0.83$ & $1.09$ & $0.95$ & $0.92$ & $-0.15$ & $ 0.27$ & $-0.13$ \\ 
  shs & $1.11$ & $0.96$ & $0.89$ & $1.11$ & $0.95$ & $0.85$ & $1.14$ & $0.97$ & $0.85$ & $1.09$ & \colBest$0.94$ & $0.93$ & $-0.11$ & $ 0.34$ & $-0.20$ \\ 
  iG & $1.08$ & $0.97$ & $0.88$ & $1.06$ & $0.95$ & \colBest$0.83$ & $1.08$ & $0.96$ & $0.83$ & $1.06$ & $0.96$ & $0.92$ & $-0.14$ & $ 0.21$ & $-0.08$ \\ 
  dhs-GP & $1.11$ & $0.95$ & $0.88$ & $1.10$ & $0.95$ & $0.84$ & $1.13$ & \colBest$0.96$ & $0.83$ & $1.09$ & $0.95$ & $0.93$ & $-0.22$ & $ 0.28$ & $-0.13$ \\ 
  shs-GP & $1.11$ & $0.96$ & $0.88$ & $1.11$ & $0.96$ & $0.84$ & $1.13$ & $0.97$ & $0.84$ & $1.09$ & $0.95$ & $0.93$ & $-0.15$ & $ 0.33$ & $-0.16$ \\ 
  iG-GP & $1.11$ & $0.96$ & $0.88$ & $1.10$ & $0.95$ & $0.84$ & $1.13$ & $0.97$ & $0.84$ & $1.09$ & $0.95$ & $0.92$ & $-0.07$ & $ 0.34$ & $-0.17$ \\ 
  dhs-GPt & $1.07$ & $0.97$ & \colBest$0.88$ & $1.06$ & \colBest$0.95$ & $0.83$ & $1.08$ & $0.96$ & \colBest$0.83$ & $1.06$ & $0.96$ & $0.92$ & $-0.15$ & $ 0.29$ & $-0.08$ \\ 
  shs-GPt & $1.12$ & $1.01$ & $0.91$ & $1.12$ & $1.04$ & $0.89$ & $1.17$ & $1.05$ & $0.90$ & $1.06$ & $0.97$ & \colBest$0.90$ & $-0.13$ & $ 0.35$ & $-0.03$ \\ 
  iG-GPt & $1.11$ & $0.96$ & $0.88$ & $1.10$ & $0.96$ & $0.84$ & $1.13$ & $0.97$ & $0.83$ & $1.08$ & $0.95$ & $0.92$ & $-0.19$ & $ 0.26$ & $-0.11$ \\ 
   \bottomrule
\end{tabular*}
\begin{tablenotes}[para,flushleft]
\scriptsize{\textit{Notes}: For model abbreviations, see Section \ref{subsec:UCQR} and Tab. \ref{tab:models}. Quantile weighted continuous ranked probability scores (CRPS) for different weighting schemes. The benchmark shows actual values shaded in light grey, all other models are shown relative to the benchmark. The best performing model is indicated in dark grey shading.}
\end{tablenotes}
\end{threeparttable}
\end{scriptsize}
\end{center}
\caption{Density forecast metrics for EA inflation.} 
\label{tab:EA-dens}
\end{table*}

Tables \ref{tab:US-dens} to \ref{tab:EA-dens} report the quantile weighted CRPS variants and LPS for the US, UK and EA as averages over the holdout. They are benchmarked relative to the UC-SV model (as ratios for CRPSs, and differences for average LPSs). The row for the benchmark displays actual metrics, all other models are shown relative to the benchmark. For CRPSs, smaller numbers are better and values below one indicate that the respective model improves upon the benchmark. By contrast, larger numbers for LPSs signal improvements and vice versa.

Investigating the results in detail, UC-SV and UC-SVM appear to be tough benchmarks to improve upon for all three considered economies. UC-SVM typically performs slightly better than the conventional UC-SV. These results confirm the findings of \citet{chan2017stochastic} and \citet{huber2020dynamic}. The conditional-mean based models are particularly strong for short-horizon forecasts, with most favorable metrics in all parts of the distribution targeted by the quantile weighted CRPSs. In fact, the best performing specification for one-quarter ahead forecasts in terms of CRPSs is UC-SVM in almost all cases across the three economies.

For multi-step ahead forecasts, the additional distributional flexibility for the quantile-based model tends to pay off in terms of predictive accuracy. This finding is especially pronounced at the three-year ahead horizon. One of the UCQR specifications is identified as the best performing model measured with the CRPS variants for all considered higher-order forecasts, with the exception of one-year ahead forecasts for the US. The quantile weighted versions of CRPSs target different parts of the distribution, and thus allow to investigate where overall gains stem from in more detail. The results indicate that UCQR, again for higher-order forecasts, results in pronounced gains, especially when focusing on either the left, right or both tails via the weighting scheme.

Within the class of quantile-based models, some differences with respect to the considered economy are apparent. Allowing for conditional heteroskedasticity (TVS) within quantiles results in more accurate forecasts for the US and the EA. In these two economies, it is worth mentioning that the homoskedastic variant (TIS) fails to improve upon the mean-based models in many cases, while TVS shows improvements ranging between $10$ to $20$ percent lower CRPSs depending on the economy, horizon and weighting scheme. This is different in the UK, where TIS seems to be better suited to model the predictive distribution of inflation. However, these gains are more modest, and in general, the differences in forecast metrics are smaller for the UK than the US or EA. Interestingly, LPSs and CRPSs do not necessarily agree on model selection. One of the UCQR-models is still identified as the best performing specification, but by the LPS metric, it is usually one of the homoskedastic QRs.

Differences in predictive accuracy between prior specifications for UCQR are often muted. In many cases one of the two dynamic shrinkage priors is identified as the best performing model. In cases where the \texttt{iG} prior works best, it does so typically at a small margin. This corroborates the findings in \citet{huber2020dynamic}, who show that imposing shrinkage in small-scale time-varying parameter models never severely hurts predictive accuracy, but in many instances yields improvements. It is worth mentioning that the dynamic horseshoe (\texttt{dhs}) prior performs well most consistently on average compared to the static horseshoe (\texttt{shs}). Turning to the consequences of post-processing quantile estimates to achieve a noncrossing quantile solution, I find that using a fixed bandwidth in the GPR usually yields similar results compared to those allowing for time-variation in this parameter. Interestingly, the raw quantile estimates which are not guaranteed to be monotonic exhibit the best forecast performance in a small number of cases. In a nutshell, adjusting quantiles ex post neither helps nor hurts predictive accuracy much.

Summarizing, four key findings are worth noting. First, variants of UCQR perform particularly well for higher-order and tail forecasts. For one-quarter ahead forecasts, the mean-based competing models often indicate better performance, but only at small margins and UCQR is competitive. Second, there is some heterogeneity with respect to the considered economies. Allowing for conditional heteroskedasticity within quantiles pays off for the US and EA, while homoskedastic scales are sufficient for the case of the UK. Third, imposing dynamic shrinkage helps predictive accuracy in many cases. The dynamic horseshoe exhibits the most consistent improvements across forecast metrics, horizons and economies. Finally, adjusting for noncrossing quantiles has only minor implications with respect to predictive accuracy.

\subsection{Higher-order moments of predictive distributions}
To shed light on which features in the resulting predictive distributions yield gains in tail forecast performance, I investigate their shapes in more detail in the following. Note that the kernel smoothing of the estimated quantiles used to compute LPSs can also be employed to generate random samples from the estimated predictive distributions. 

In order to compute and investigate higher-order moments of these distributions, I focus on one-year ahead forecasts for UC-SV, UC-SVM, UCQR-TIS-GPt-dhs and UCQR-TVS-GPt-dhs, and use the following procedure. First, I generate a sample of $3,000$ random numbers from the respective predictive distribution which I then use to compute the mean, variance, excess kurtosis and skewness of this sample. This step is iterated for $1,000$ times. Second, to obtain numerical standard errors for the estimates of these moments using a type of bootstrap-procedure, I compute the $5$th and $95$th percentile over these $1,000$ replicates.

\begin{figure}[!ht]
  \begin{subfigure}{\textwidth}
  \caption{US}
  \includegraphics[width=\textwidth]{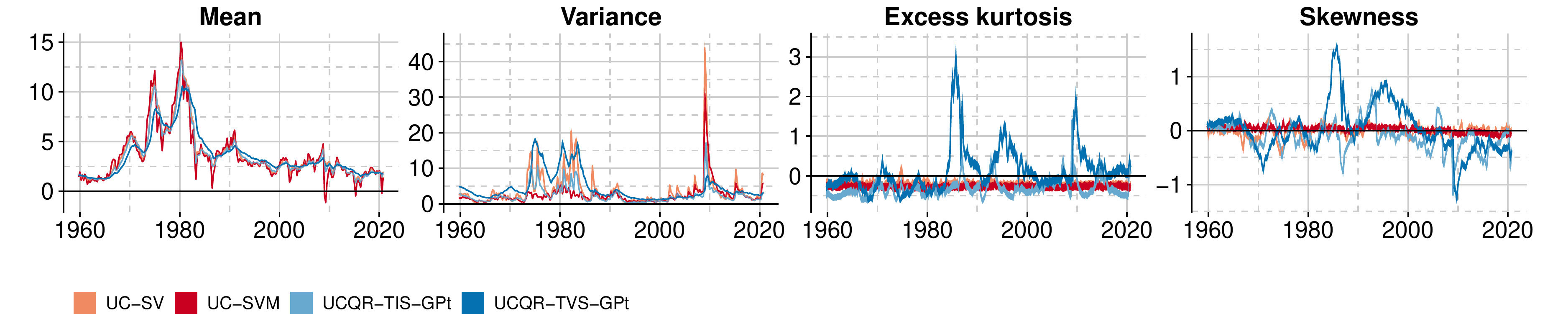}
  \end{subfigure}
  \begin{subfigure}{\textwidth}
  \caption{UK}
  \includegraphics[width=\textwidth]{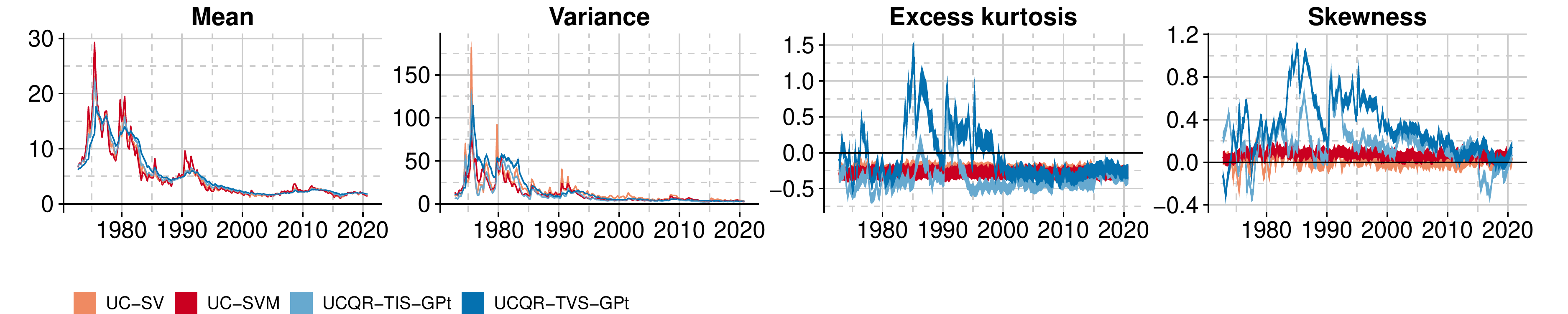}
  \end{subfigure}
  \begin{subfigure}{\textwidth}
  \caption{EA}
  \includegraphics[width=\textwidth]{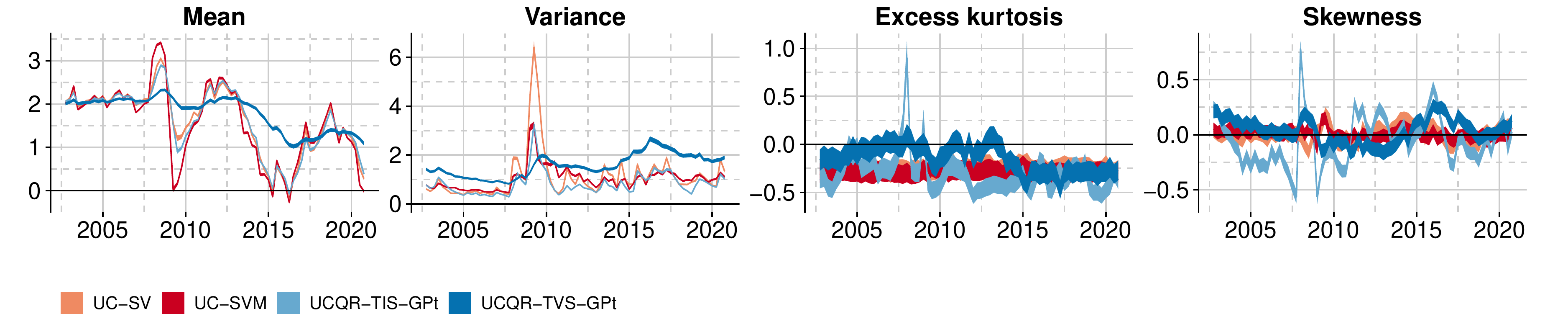}
  \end{subfigure}
  \caption{Higher-order moments of the one-year ahead predictive distributions.}\label{fig:moments}\vspace*{-1em}
  \caption*{\scriptsize{\textit{Notes:} Unobserved component model with stochastic volatility (UC-SV), stochastic volatility in mean (UC-SVM) and unobserved component quantile regression (UCQR) with time-invariant (TIS) or time-varying (TVS) scale parameter, adjusted using a Gaussian process (GP) regression with time-varying bandwidth. All models are estimated using the dynamic horseshoe prior. The lines mark numerical $5$th and $95$th percentiles across $1,000$ replicates.}}
\end{figure}

The resulting empirical moments are shown in Figure \ref{fig:moments}. Starting with the mean of the predictive distribution, several stylized facts across all three economies can be established. While UC-SV and UC-SVM often show similar paths, UC-SVM typically produces more pronounced high-frequency movements in the context of large-variance shocks. This feature is due to feedback effects between inflation and its volatility. For the US, \citet{chan2017stochastic} shows that the level-volatility relationship was positive pre-$1980$, but turned insignificant or negative afterwards. In other words, large-variance shocks coincided with higher inflation early in the sample, while large variance shocks later in the sample relate to downward movements of inflation. 

Comparing the two mean-based competing models to the QRs, predictive means from UCQR-TIS often exhibits a similar path, while shifts in the location of the predictive distribution in the case of UCQR-TVS often occur with a delay and are less pronounced. This finding is similar to in-sample evidence, and overall, the predictive mean for the latter model displays a less noisy and smoother evolution. The more aggressive fit of the mean-based models, in light of the forecast results, yields gains in predictive accuracy in the short horizon. However, it appears that a less aggressive fit tends to pay off for higher-order forecasts.

Turning to the variances, several findings are worth noting. In general, time-varying variances for the US and UK are clearly featured for all competing models. This requirement is mostly dictated by the comparatively long sample vis-\`{a}-vis the EA spanning different economic phases \citep[see, e.g.,][]{clark2011real}. By contrast, variances in predictive distributions for the EA are comparatively stable, which corroborates findings by \citet{jarocinski2018inflation} and \citet{huber2020multi}, who detect only limited evidence in favor of stochastic volatility for inflation in the EA. Note that this finding suggests that models for the conditional mean could safely neglect time-varying volatilities, while conditional heteroskedasticity within quantiles via TVS receives empirical support. The UC-SVM model overall yields the narrowest predictive distributions, apart from during the Great Recession in the US. This may be explained by the notion that the volatility process is featured as a predictor, providing a better fit, thereby also reducing the predictive variance. Interestingly, for the US and the EA, predictive variances in normal economic times are typically slightly larger for UCQR-TVS compared to the benchmark models, which can be related to the less agressive fit with respect to the location of the distribution. During the Great Recession, or the Covid-19 pandemic for the case of the US, however, the predictive distributions are narrower than those of the mean-based models or UCQR-TIS.

Turning to the higher-order moments -- excess kurtosis and skewness -- differences between model specifications are more striking. Lower variances during recessionary episodes may in part be explained by noting that particularly UCQR-TVS signals periods of substantial excess kurtosis, pointing to heavier than normal tails of the predictive distribution. Moreover, the resulting distributions are skewed (positively and negatively) during different parts of the samples across economies. Interestingly, UCQR-TIS usually exhibits excess kurtosis values around zero, and less pronouncedly skewed predictive distributions, indicating that this feature arises from allowing for conditional heteroskedasticity within quantiles. Given the comparatively worse performance of UCQR-TVS in the case of the UK, the forecast results suggest that such features are irrelevant to improve density forecast accuracy, while they yield gains for the US and the EA.

It is worth discussing several periods in more detail. Strikingly, excess kurtosis in predictive distribution for the US occurs after the high-inflation period of the late $1970$s. The skewness parameter in conjunction with these heavier than normal tails in the mid $1980$s suggests substantial upward inflation risk during the Volcker chairmenship. Later in the sample, however, during the global financial crisis, non-zero excess kurtosis is again measured, but with a negatively skewed distribution, pointing to disinflationary pressures and downward risk. This notion may be linked to the findings of a sign-switch in the relationship between inflation and its volatility identified in \citet{chan2017stochastic}. By contrast, the skewness parameter for UCQR-TIS in the UK is positive for most of the sample, pointing mainly towards upward risk early in the sample which has declined since the early $1990$s. For the EA, some periods of skewed distributions are featured, particularly during the European debt crisis, but they are not as pronounced as those for the other economies.

Summarizing, this analysis indicates that forecast gains for the UCQR models can in part be explained by allowing for more flexible predictive distributions featuring both heavier than normal tails and skewness. While such features improve longer-horizon and tail forecasts in the US and the EA, this is not necessarily the case for the UK.

\subsection{Tail risks in predictive distributions}
\begin{figure}[!ht]
  \begin{subfigure}{\textwidth}
  \caption{US}
  \includegraphics[width=\textwidth]{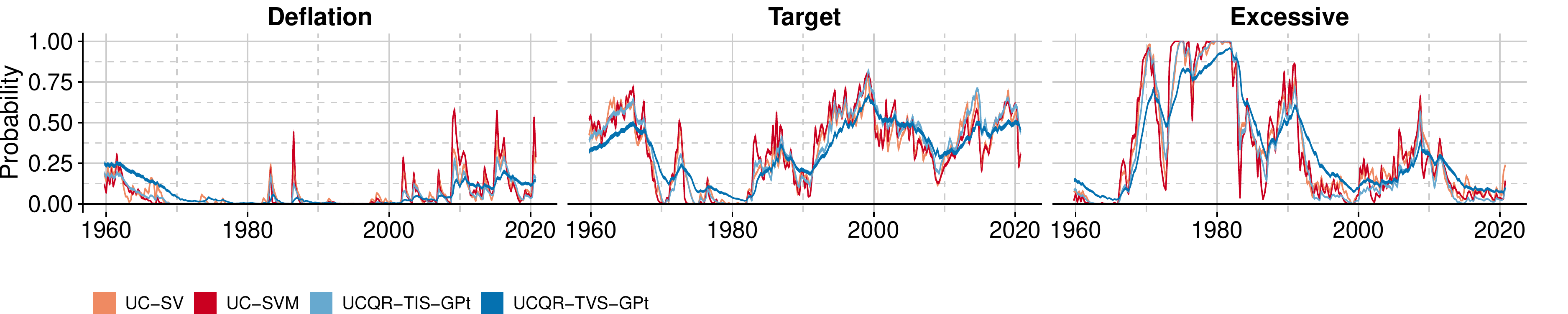}
  \end{subfigure}
  \begin{subfigure}{\textwidth}
  \caption{UK}
  \includegraphics[width=\textwidth]{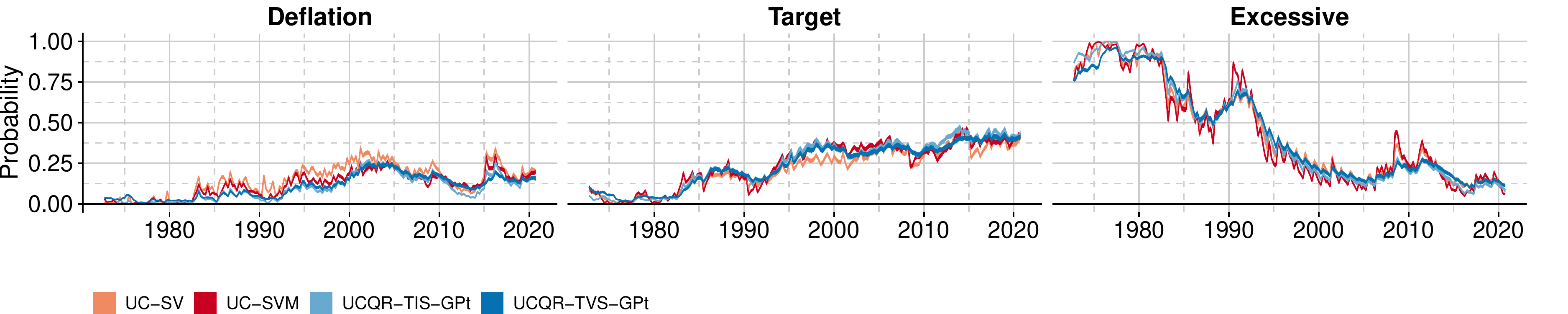}
  \end{subfigure}
  \begin{subfigure}{\textwidth}
  \caption{EA}
  \includegraphics[width=\textwidth]{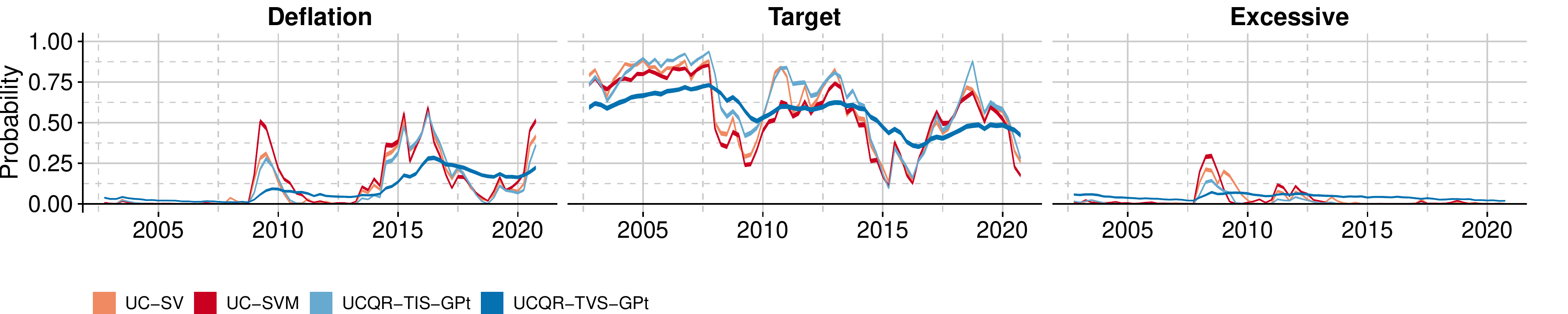}
  \end{subfigure}
  \caption{Predictive one-year ahead scenario probabilities.}\label{fig:probs}\vspace*{-1em}
  \caption*{\scriptsize{\textit{Notes:} Unobserved component model with stochastic volatility (UC-SV), stochastic volatility in mean (UC-SVM) and unobserved component quantile regression (UCQR) with time-invariant (TIS) or time-varying (TVS) scale parameter, adjusted using a Gaussian process (GP) regression with time-varying bandwidth. All models are estimated using the dynamic horseshoe prior. Scenarios (a) \textit{Deflation} with $\Pr(\pi_{t+h} < 0)$, (b) \textit{Target} with $\Pr(1 \leq \pi_{t+h} \leq 3)$, and (c) \textit{Excessive} with $\Pr(\pi_{t+h} > 4)$. The lines mark numerical $5$th and $95$th percentiles across $1,000$ replicates.}}
\end{figure}

To complement this discussion of tail risks in inflation, and to quantify both upside and downside risk, I compute the probabilities of three different inflation-scenarios: (a) \textit{Deflation} with $\Pr(\pi_{t+h} < 0)$, (b) \textit{Target} with $\Pr(1 \leq \pi_{t+h} \leq 3)$, and (c) \textit{Excessive} with $\Pr(\pi_{t+h} > 4)$. An interesting aspect of this exercise is that while increased variances for UC-SV and UC-SVM usually widen the predictive distribution symetrically around the mean, this is not necessarily the case for UCQR, which might affect the evolution of scenario probabilities. Moreover, as suggested above, there are several differences between estimates of the predictive mean, variance, excess kurtosis and skewness. All of these may affect the probability of inflation lying within the specified bounds. I again use $1,000$ replicates of samples from the predictive distributions and compute the numerical $5$th and $95$th percentiles. The results are shown in Figure \ref{fig:probs}.

The previous findings of UCQR-TVS producing smoother moments of predictive distributions carry over to this exercise. Overall, all models tend to agree on the respective probabilities of the considered scenarios, with some differences in the case of the EA. For the US, deflationary risks were virtually non-existent between $1970$ and $1985$. They appear to be a threat particularly in more recent years, with similarities to the case of the EA. Conversely, probabilities close to one of excessive inflation are visible during the high-inflation periods during the $1980$s for the US and the UK, but show overall declines with minor peaks for the rest of the sample, while being only minor in the EA. This is consistent with the overall trend towards more stable prices in the last three decades. Notable upward risks occur prior to the early $1990$s recession in the US and the ERM crisis in $1992$ in the UK; moreover, I detect inflationary risks just before the Great Recession for both economies, and also for the EA.

Turning to the \textit{Target} scenario, which reflects the ability of central banks to keep inflation anchored at the respective target, I find several interesting patterns. Both in the US and the UK, there is a trend towards increasing probabilities of keeping inflation within the interval between one and three percent since the $1980$s and the $1990$s, respectively. A theoretical argument for this break is provided by noting that the Volcker disinflation in the case of the Federal Reserve, and the adoption of inflation targeting by the Bank of England in $1992$, constitutes a policy regime change in the approach to central banking. For a theoretical model discussing breaking the link between inflation and inflation volatility via policy regime changes, and, with it, stabilization of price dynamics, see \citet{cukierman1986theory}.

Further increasing probabilities of keeping inflation on target in recent years may be linked to more transparent central bank communication and expectation management strategies \citep[see, e.g.,][]{blinder2008central}. Compared to the Federal Reserve and the Bank of England, the European Central Bank exhibits stable and higher probabilities of fulfilling its target measured by UCQR (albeit for a much shorter sampling period). It is worth mentioning that a slight downward trend in target-probabilities is observable for the EA particularly in the aftermath of economic crises, coinciding with upward movements in deflationary pressures.

\section{Closing remarks}\label{sec:conclusions}
In this paper, I discuss methods for estimating Bayesian time-varying parameter quantile regressions featuring conditional heteroskedasticity. I rely on data augmentation schemes to render the model conditionally Gaussian and develop an efficient Gibbs sampling algorithm. The high-dimensional parameter space is regularized via dynamic shrinkage priors. 

A simple variant of the resulting framework is applied to modeling the distribution of inflation in the United States, the United Kingdom and the euro area. This model is a quantile-based version of the popular unobserved component model, which has had great success in modeling and forecasting inflation. An out-of-sample forecast horserace indicates that the resulting models are competitive and perform particularly well for higher-order and tail forecasts. A detailed analysis of the resulting predictive distributions reveals that they are sometimes skewed and feature heavy tails. Investigating implied probabilities of several inflation scenarios reveals that excessive inflation was a threat early in the sample in the US and the UK. Deflationary risks emerge mostly in more recent years.

\small{\setstretch{0.85}
\addcontentsline{toc}{section}{References}
\bibliographystyle{custom.bst}
\bibliography{lit}}\normalsize\clearpage

\begin{appendices}\crefalias{section}{appsec}
\begin{center}
{\Large\sffamily\textbf{Appendix}}
\end{center}

\setcounter{equation}{0}
\renewcommand\theequation{A.\arabic{equation}}
\section{MCMC algorithm}\label{app:A:Sampling}
\subsection{Sampling auxiliary variables}
The likelihood based on (\ref{eq:QR-N}) allows for constructing an efficient Gibbs sampling algorithm, as discussed in \citet{kozumi2011gibbs}. In particular, the conditional posterior distribution (with $\bullet$ denoting conditioning on all other parameters of the model and the data) for the auxiliary variable $v_{pt}$ is given by:
\begin{equation}
v_{pt}|\bullet\sim\mathcal{GIG}\left(\frac{1}{2},\frac{(y_t - \bm{x}_t'\bm{\beta}_{pt})^2}{\tau_p^2\sigma_{pt}},\frac{2}{\sigma_{pt}} + \frac{\theta_p^2}{\tau_p^2\sigma_{pt}}\right),\label{eq:postauxiliary}
\end{equation}
where $\mathcal{GIG}$ is the generalized inverse Gaussian distribution.\footnote{The generalized inverse Gaussian (GIG) distribution is parameterized such that a random variable $X\sim\mathcal{GIG}(\lambda,\xi,\psi)$ with probability density function $f(x;\lambda,\xi,\psi)=x^{\lambda-1}\exp\left(-(\xi/x + \psi x)/2\right)$.} 

\subsection{Sampling the scale parameter of the asymmetric Laplace distribution}\label{app:ALDTVP}
The time-varying scale parameter $\sigma_{pt}$ can be sampled using a random walk Metropolis-Hastings algorithm similar to \citet{jacquier2002bayesian} for the stochastic volatility model. It is worth mentioning that the approximation of the AL$_p$ distribution moves the scale parameter to the mean of the model, which implies that this algorithm shares commonalities with samplers of stochastic volatility in mean models.

For notational simplicity, define $h_{pt} = \log(\sigma_{pt})$. The prior on the initial state is $h_{p0}\sim\mathcal{N}(m_{0},\varsigma_0^2)$, with $m_{0}=0$ and $\varsigma_0^2=1$ in the empirical application. The conditional posterior of the initial state under this prior is given by $h_{p0}\sim\mathcal{N}(\bar{m}_0,\bar{S}_0)$ and corresponding moments $\bar{m}_0 = \varsigma_0^2(m_0/\varsigma_0^2 + h_{p1}/\varsigma_p^2)$ and $\bar{S}_0 = (\varsigma_0^2\varsigma_p^2)/(\varsigma_0^2 + \varsigma_p^2)$. The state equation (\ref{eq:scalestate-eq}) defines the conditional priors at each point in time $t$:
\begin{itemize}[align=left]
\item[---] For time $t=1,\hdots,T-1$, it is given by $h_{pt}\sim\mathcal{N}(\bar{m}_{pt},\bar{S}_{pt})$ where $\bar{m}_{pt} = (h_{pt-1} + h_{pt+1})/2$ and $\bar{S}_{pt} = \varsigma_p^2/2$.
\item[---] For the final value at $t=T$, the conditional prior density is $h_{pT}^{(r)}\sim\mathcal{N}(\bar{m}_{pT},\bar{S}_{pT})$ with $\bar{m}_{pT} = h_{pT-1}$ and $\bar{S}_{pT} = \varsigma_p^2$.
\end{itemize}

Let $r$ refer to the currently accepted draw of the respective quantity and define $z_{pt} = v_{pt}/\sigma_{pt}^{(r)}$, see \citet{kozumi2011gibbs} for details. Combining the conditional priors with the likelihood defined in (\ref{eq:QR-N}), a simple accept/reject sampling algorithm with acceptance probability $\zeta_{pt}$ on a $t$-by-$t$ basis can be derived using a random walk proposal, $h_{pt}^{(\ast)}\sim\mathcal{N}(h_{pt}^{(r)},c)$ with $c$ being a tuning parameter:
\begin{equation*}
\zeta_{pt} = \min\left(\frac{\mathcal{N}\left(y_{t};\bm{x}_t'\bm{\beta}_{pt} + \theta_p z_{pt}\exp(h_{pt}^{(\ast)}),\tau_{p}^2 z_{pt}\exp(h_{pt}^{(\ast)})^2\right)\times\mathcal{N}(\exp(h_{pt}^{(\ast)});\bar{m}_{pt},\bar{S}_{pt})}{\mathcal{N}\left(y_{t};\bm{x}_t'\bm{\beta}_{pt} + \theta_p z_{pt}\exp(h_{pt}^{(r)}),\tau_{p}^2 z_{pt}\exp(h_{pt}^{(r)})^2\right)\times\mathcal{N}(\exp(h_{pt}^{(r)});\bar{m}_{pt},\bar{S}_{pt})},1\right).
\end{equation*}

The proposed value $h_{pt}^{(\ast)}$ is accepted with probability $\zeta_{pt}$, otherwise, the previous draw $h_{pt}^{(r)}$ is retained. After obtaining the full history of $\{h_{pt}\}_{t=1}^T$, the state innovation variance $\varsigma_p^2$ can be sampled from its inverse Gamma conditional posterior distribution using standard moments for Bayesian linear regression models. 

The conditional posterior for a time-invariant scale parameter $\sigma_p$ is:
\begin{equation}
\sigma_p|\bullet\sim\mathcal{G}^{-1}\left(\frac{a + 3T}{2},\frac{b + 2\sum_{t=1}^T v_{pt}}{2} + \frac{\sum_{t=1}^T (y_t - \bm{x}_t'\bm{\beta}_{pt} - \theta_p v_{pt})^2}{2\tau_p^2 v_{pt}}\right).\label{eq:postscale}
\end{equation}

\subsection{Sampling the time-varying parameters and shrinkage processes}
Given the state innovation variances in $\bm{\Omega}_{pt}$, conventional Kalman-filter based methods such as forward-filtering backward-sampling \citep[FFBS, see][]{carter1994gibbs,fruhwirth1994data} or faster alternatives \citep[see, e.g.,][]{chan2009efficient,hauzenberger2019fast} can be used to draw the full history of the time-varying, quantile-specific regression coefficients $\{\bm{\beta}_{pt}\}_{t=1}^T$.

Time-variation in the state innovation variances of the regression coefficients, collected in $\bm{\Omega}_{pt}$, is governed by (\ref{eq:shrinkstate}). Given $\bm{\beta}_{pt}$, a mixture representation of the Z-distribution using P\'olya-Gamma (denoted by $\mathcal{PG}$) random variables can be employed:
\begin{equation*}
\nu_{pk,t}|\xi_{pk,t}\sim\mathcal{N}(\xi_{pk,t}^{-1}(c-d)/2,\xi_{pk,t}^{-1}), \quad \xi_{pk,t}\sim\mathcal{PG}(c+d,0),
\end{equation*}
for $k=1,\hdots,K$. The approximation renders (\ref{eq:shrinkstate}) conditionally Gaussian, and an appropriate algorithm can again be used to sample the full history of the state innovation variances. This procedure is similiar to popular approaches in stochastic volatility models \citep[see, e.g.,][]{kim1998stochastic}, and described in detail in \citet{kowal2019dynamic}.

For the \texttt{iG} prior, posterior moments of the state innovation variances have a standard textbook form and are not reproduced here; corresponding moments for the \texttt{shs} prior can be found in \citet{makalic2015simple} and \citet{huber2020dynamic}.

\subsection{The full sampling algorithm}
After initializing the sampler, the Markov chain Monte Carlo (MCMC) algorithm iterates through the following steps:
\begin{enumerate}[align=left,label=(\arabic*)]
  \item Draw the full history of the time-varying, quantile-specific regression coefficients $\{\bm{\beta}_{pt}\}_{t=1}^T$ conditional on all other parameters using FFBS based on the observation and state equations given by (\ref{eq:measurement}) and (\ref{eq:state-eq}).
  \item Using the mixture representation of the Z-distribution and conditional on $\{\bm{\beta}_{pt}\}_{t=1}^T$, draw the full history of the state innovation variances $\{\omega_{pk,t}\}_{t=1}^T$ for $k=1,\hdots,K$ again by FFBS in the case of the dynamic horseshoe. For the static horseshoe and inverse Gamma priors, the conditional posteriors have closed-form solutions and are discussed above. Subsequently, the regression parameters in (\ref{eq:shrinkstate}) and auxiliary quantities for the mixture representation can be drawn from their conditional posterior distributions \citep[see][for details]{kowal2019dynamic}.
  \item Conditional on all other parameters of the model, the auxiliary variables $v_{pt}$ can be drawn from their generalized inverse Gaussian-distributed conditional posterior period-by-period with moments given in (\ref{eq:postauxiliary}).
  \item The full history of the scale parameter $\{\sigma_{pt}\}_{t=1}^T$, conditional on all other parameters of the model, can be sampled using a random walk Metropolis-Hastings algorithm as in \citet{jacquier2002bayesian} which is discussed in Section \ref{app:ALDTVP} of this Appendix. The likelihood is given by (\ref{eq:QR-N}), while the law of motion given by (\ref{eq:scalestate-eq}) defines the conditional prior at each point in time. In the time-invariant case, it is sampled from its inverse Gamma conditional posterior with moments provided in (\ref{eq:postscale}).
  \item Forecasts for the dynamic shrinkage process and the TVPs can be obtained via simulation based methods. For multi-step ahead forecasts, ony may specify the vector $\bm{x}_t$ accordingly and produce direct forecasts. 
\end{enumerate}
The produced MCMC output can be used in the context of the GPR discussed in Section \ref{subsec:noncrossing} to post-process raw estimates for achieving noncrossing quantile estimates.

\section{Additional forecasting results}\label{app:B:Results}
\subsection{Quantile scores}
Tables \ref{tab:US-QS} to \ref{tab:EA-QS} contain additional forecast results on point forecast performance by quantile. Note that the QS for $p=0.5$ correspond to the mean absolute error (MAE) and is thus a conventional point forecast metric in the case of conditional-mean based models.

\begin{table*}[ht]
\begin{center}
\begin{scriptsize}
\begin{threeparttable}
\begin{tabular*}{\textwidth}{@{\extracolsep{\fill}} lrrrrrrrrrrrrrrr}
  \toprule
 & \multicolumn{15}{c}{\textbf{QS}} \\
 \cmidrule(l){2-16}
\multicolumn{1}{r}{$p=$} & \multicolumn{3}{c}{$0.05$} & \multicolumn{3}{c}{$0.1$} & \multicolumn{3}{c}{$0.5$ (MAE)} & \multicolumn{3}{c}{$0.9$} & \multicolumn{3}{c}{$0.95$} \\
 \cmidrule(l){2-4}\cmidrule(l){5-7}\cmidrule(l){8-10}\cmidrule(l){11-13}\cmidrule(l){14-16}
\multicolumn{1}{r}{$h=$} & 1 & 4 & 12 & 1 & 4 & 12 & 1 & 4 & 12 & 1 & 4 & 12 & 1 & 4 & 12 \\
  \midrule
UC-SV & \colBench$0.53$ & \colBench$0.61$ & \colBench$0.81$ & \colBench$0.78$ & \colBench$0.93$ & \colBench$1.19$ & \colBench$1.33$ & \colBench$1.76$ & \colBench$2.16$ & \colBench$0.59$ & \colBench$0.93$ & \colBench$1.31$ & \colBench$0.37$ & \colBench$0.63$ & \colBench$0.95$ \\ 
  UC-SVM & $1.04$ & $1.17$ & $1.27$ & $1.00$ & $1.07$ & $1.16$ & \colBest$0.97$ & \colBest$0.97$ & $1.04$ & \colBest$0.89$ & $0.94$ & $1.08$ & \colBest$0.83$ & $0.93$ & $1.15$ \\ 
  \midrule
  \multicolumn{16}{c}{\textbf{UCQR-TIS}} \\
  \midrule
    dhs & $1.08$ & $1.08$ & $1.11$ & $1.01$ & $1.01$ & $1.01$ & $1.04$ & $1.00$ & $0.98$ & $1.01$ & $1.04$ & $1.05$ & $1.06$ & $1.15$ & $1.17$ \\ 
  shs & $1.13$ & $1.24$ & $1.27$ & $1.04$ & $1.08$ & $1.09$ & $1.00$ & $0.99$ & $1.00$ & $0.98$ & $1.01$ & $1.02$ & $1.04$ & $1.11$ & $1.11$ \\ 
  iG & $1.23$ & $1.54$ & $1.56$ & $1.07$ & $1.24$ & $1.25$ & $0.99$ & $0.99$ & $1.04$ & $1.07$ & $1.10$ & $1.15$ & $1.30$ & $1.38$ & $1.41$ \\ 
  dhs-GP & $1.08$ & $1.11$ & $1.13$ & $1.02$ & $1.01$ & $1.00$ & $1.04$ & $1.00$ & $0.98$ & $1.00$ & $1.04$ & $1.05$ & $1.06$ & $1.16$ & $1.18$ \\ 
  shs-GP & $1.12$ & $1.22$ & $1.25$ & $1.04$ & $1.08$ & $1.09$ & $1.00$ & $0.99$ & $1.00$ & $0.98$ & $1.00$ & $1.03$ & $1.02$ & $1.10$ & $1.13$ \\ 
  iG-GP & $1.11$ & $1.20$ & $1.23$ & $1.04$ & $1.09$ & $1.10$ & $1.00$ & $0.99$ & $1.00$ & $0.98$ & $1.01$ & $1.03$ & $1.02$ & $1.08$ & $1.11$ \\ 
  dhs-GPt & $1.22$ & $1.53$ & $1.52$ & $1.09$ & $1.25$ & $1.27$ & $0.99$ & $0.99$ & $1.05$ & $1.11$ & $1.13$ & $1.17$ & $1.26$ & $1.32$ & $1.37$ \\ 
  shs-GPt & $1.22$ & $1.54$ & $1.56$ & $1.06$ & $1.22$ & $1.26$ & $1.00$ & $1.00$ & $1.04$ & $1.07$ & $1.10$ & $1.15$ & $1.30$ & $1.35$ & $1.39$ \\ 
  iG-GPt & $1.08$ & $1.11$ & $1.13$ & $1.02$ & $1.00$ & $1.00$ & $1.04$ & $1.00$ & $0.98$ & $1.00$ & $1.04$ & $1.05$ & $1.06$ & $1.17$ & $1.19$ \\ 
  \midrule
  \multicolumn{16}{c}{\textbf{UCQR-TVS}} \\
  \midrule
  dhs & $1.07$ & $0.95$ & $0.74$ & $1.06$ & $0.91$ & \colBest$0.81$ & $1.16$ & $1.01$ & \colBest$0.95$ & $1.25$ & $1.07$ & $0.92$ & $1.28$ & $1.05$ & $0.83$ \\ 
  shs & \colBest$0.98$ & $0.95$ & $0.79$ & \colBest$0.99$ & $0.90$ & $0.84$ & $1.17$ & $1.04$ & $0.98$ & $1.15$ & $1.03$ & $0.92$ & $1.16$ & $0.93$ & $0.74$ \\ 
  iG & $1.03$ & $0.92$ & $0.75$ & $1.01$ & $0.89$ & $0.83$ & $1.12$ & $1.00$ & $0.97$ & $1.10$ & $0.95$ & $0.84$ & $1.16$ & \colBest$0.89$ & $0.65$ \\ 
  dhs-GP & $1.06$ & $0.95$ & \colBest$0.73$ & $1.05$ & $0.91$ & $0.81$ & $1.16$ & $1.02$ & $0.96$ & $1.27$ & $1.09$ & $0.93$ & $1.26$ & $1.05$ & $0.82$ \\ 
  shs-GP & $1.05$ & $0.94$ & $0.74$ & $1.04$ & $0.90$ & $0.83$ & $1.15$ & $1.02$ & $0.97$ & $1.16$ & $1.03$ & $0.92$ & $1.17$ & $0.97$ & $0.79$ \\ 
  iG-GP & $1.05$ & $0.94$ & $0.74$ & $1.04$ & $0.91$ & $0.82$ & $1.14$ & $1.02$ & $0.97$ & $1.15$ & $1.02$ & $0.91$ & $1.18$ & $0.98$ & $0.79$ \\ 
  dhs-GPt & $1.02$ & $0.92$ & $0.74$ & $1.02$ & \colBest$0.89$ & $0.81$ & $1.12$ & $1.00$ & $0.97$ & $1.10$ & $0.95$ & $0.86$ & $1.17$ & $0.89$ & $0.66$ \\ 
  shs-GPt & $1.07$ & \colBest$0.90$ & $0.76$ & $1.03$ & $0.89$ & $0.83$ & $1.14$ & $1.01$ & $0.98$ & $1.10$ & \colBest$0.92$ & \colBest$0.81$ & $1.17$ & $0.89$ & \colBest$0.64$ \\ 
  iG-GPt & $1.05$ & $0.95$ & $0.77$ & $1.05$ & $0.92$ & $0.84$ & $1.18$ & $1.02$ & $0.96$ & $1.22$ & $1.11$ & $0.95$ & $1.19$ & $1.06$ & $0.86$ \\ 
   \bottomrule
\end{tabular*}
\begin{tablenotes}[para,flushleft]
\scriptsize{\textit{Notes}: For model abbreviations, see Section \ref{subsec:UCQR} and Tab. \ref{tab:models}. Quantile scores (QS) across quantiles. QS for $p=0.5$ corresponds to the mean absolute error (MAE). The benchmark shows actual values shaded in light grey, all other models are shown relative to the benchmark. The best performing model is indicated in dark grey shading.}
\end{tablenotes}
\end{threeparttable}
\end{scriptsize}
\end{center}
\caption{Quantile scores for US inflation.} 
\label{tab:US-QS}
\end{table*}

\begin{table*}[ht]
\begin{center}
\begin{scriptsize}
\begin{threeparttable}
\begin{tabular*}{\textwidth}{@{\extracolsep{\fill}} lrrrrrrrrrrrrrrr}
  \toprule
 & \multicolumn{15}{c}{\textbf{QS}} \\
 \cmidrule(l){2-16}
\multicolumn{1}{r}{$p=$} & \multicolumn{3}{c}{$0.05$} & \multicolumn{3}{c}{$0.1$} & \multicolumn{3}{c}{$0.5$ (MAE)} & \multicolumn{3}{c}{$0.9$} & \multicolumn{3}{c}{$0.95$} \\
 \cmidrule(l){2-4}\cmidrule(l){5-7}\cmidrule(l){8-10}\cmidrule(l){11-13}\cmidrule(l){14-16}
\multicolumn{1}{r}{$h=$} & 1 & 4 & 12 & 1 & 4 & 12 & 1 & 4 & 12 & 1 & 4 & 12 & 1 & 4 & 12 \\
  \midrule
UC-SV & \colBench$0.64$ & \colBench$0.62$ & \colBench$0.57$ & \colBench$1.00$ & \colBench$1.05$ & \colBench$1.05$ & \colBench$2.69$ & \colBench$2.77$ & \colBench$2.91$ & \colBench$1.45$ & \colBench$1.65$ & \colBench$1.54$ & \colBench$0.95$ & \colBench$1.14$ & \colBench$0.97$ \\ 
  UC-SVM & \colBest$0.81$ & \colBest$0.87$ & $1.25$ & \colBest$0.91$ & $0.94$ & $1.23$ & $1.02$ & $1.00$ & $1.02$ & \colBest$0.95$ & $0.97$ & $0.98$ & \colBest$0.90$ & \colBest$0.96$ & \colBest$0.96$ \\ 
  \midrule
  \multicolumn{16}{c}{\textbf{UCQR-TIS}} \\
  \midrule
   dhs & $0.86$ & $1.03$ & $1.39$ & $0.95$ & $0.95$ & $1.13$ & $1.00$ & $1.01$ & $1.00$ & $1.03$ & $1.06$ & $1.05$ & $1.08$ & $1.11$ & $1.09$ \\ 
  shs & $0.89$ & $1.03$ & $1.60$ & $0.94$ & $0.97$ & $1.21$ & $0.98$ & $0.98$ & $0.98$ & $1.03$ & $1.04$ & $1.02$ & $1.11$ & $1.12$ & $1.04$ \\ 
  iG & $1.01$ & $1.10$ & $1.90$ & $1.00$ & $0.97$ & $1.33$ & $1.00$ & $0.96$ & $0.97$ & $1.15$ & $0.97$ & $0.97$ & $1.34$ & $1.07$ & $1.09$ \\ 
  dhs-GP & $0.87$ & $1.03$ & $1.42$ & $0.95$ & $0.96$ & $1.12$ & $0.99$ & $1.02$ & $1.01$ & $1.03$ & $1.04$ & $1.05$ & $1.10$ & $1.10$ & $1.11$ \\ 
  shs-GP & $0.88$ & $1.02$ & $1.59$ & $0.94$ & $0.97$ & $1.21$ & $0.98$ & $0.98$ & $0.98$ & $1.03$ & $1.04$ & $1.02$ & $1.10$ & $1.12$ & $1.04$ \\ 
  iG-GP & $0.88$ & $1.02$ & $1.57$ & $0.94$ & $0.96$ & $1.21$ & \colBest$0.98$ & $0.98$ & $0.98$ & $1.03$ & $1.05$ & $1.02$ & $1.10$ & $1.12$ & $1.03$ \\ 
  dhs-GPt & $1.00$ & $1.06$ & $1.89$ & $1.00$ & $0.97$ & $1.38$ & $1.00$ & $0.96$ & $0.97$ & $1.22$ & \colBest$0.97$ & \colBest$0.95$ & $1.37$ & $1.07$ & $1.06$ \\ 
  shs-GPt & $1.00$ & $1.09$ & $1.73$ & $1.00$ & $0.98$ & $1.30$ & $1.02$ & \colBest$0.96$ & $0.98$ & $1.14$ & $0.98$ & $0.97$ & $1.30$ & $1.12$ & $1.11$ \\ 
  iG-GPt & $0.87$ & $1.02$ & $1.41$ & $0.95$ & $0.96$ & $1.12$ & $1.00$ & $1.02$ & $1.01$ & $1.03$ & $1.04$ & $1.05$ & $1.11$ & $1.11$ & $1.11$ \\ 
  \midrule
  \multicolumn{16}{c}{\textbf{UCQR-TVS}} \\
  \midrule
  dhs & $0.87$ & $0.97$ & $1.14$ & $0.97$ & $0.94$ & $1.05$ & $1.01$ & $1.04$ & $0.98$ & $1.10$ & $1.14$ & $1.09$ & $1.17$ & $1.19$ & $1.16$ \\ 
  shs & $0.82$ & $0.94$ & \colBest$1.00$ & $0.94$ & $0.94$ & $1.03$ & $1.00$ & $1.02$ & $0.98$ & $1.04$ & $1.02$ & $1.00$ & $1.08$ & $1.01$ & $0.99$ \\ 
  iG & $0.82$ & $0.93$ & $1.10$ & $0.94$ & $0.93$ & $1.07$ & $1.00$ & $1.00$ & $0.96$ & $1.06$ & $1.08$ & $1.01$ & $1.10$ & $1.08$ & $1.01$ \\ 
  dhs-GP & $0.87$ & $0.96$ & $1.14$ & $0.97$ & $0.95$ & $1.06$ & $1.01$ & $1.04$ & $0.98$ & $1.11$ & $1.15$ & $1.11$ & $1.16$ & $1.18$ & $1.14$ \\ 
  shs-GP & $0.85$ & $0.94$ & $1.15$ & $0.96$ & $0.95$ & $1.06$ & $1.00$ & $1.02$ & $0.97$ & $1.07$ & $1.09$ & $1.06$ & $1.11$ & $1.11$ & $1.04$ \\ 
  iG-GP & $0.85$ & $0.95$ & $1.15$ & $0.95$ & $0.94$ & $1.04$ & $1.00$ & $1.01$ & $0.96$ & $1.07$ & $1.09$ & $1.05$ & $1.12$ & $1.11$ & $1.04$ \\ 
  dhs-GPt & $0.82$ & $0.93$ & $1.11$ & $0.93$ & \colBest$0.93$ & $1.05$ & $1.00$ & $1.00$ & \colBest$0.95$ & $1.05$ & $1.08$ & $1.02$ & $1.10$ & $1.09$ & $1.01$ \\ 
  shs-GPt & $0.84$ & $0.94$ & $1.03$ & $0.93$ & $0.94$ & $1.05$ & $1.00$ & $1.00$ & $0.97$ & $1.01$ & $1.00$ & $0.98$ & $1.03$ & $0.99$ & $0.97$ \\ 
  iG-GPt & $0.88$ & $0.99$ & $1.19$ & $0.96$ & $0.95$ & $1.07$ & $1.01$ & $1.03$ & $0.98$ & $1.14$ & $1.16$ & $1.12$ & $1.20$ & $1.21$ & $1.16$ \\ 
   \bottomrule
\end{tabular*}
\begin{tablenotes}[para,flushleft]
\scriptsize{\textit{Notes}: For model abbreviations, see Section \ref{subsec:UCQR} and Tab. \ref{tab:models}. Quantile scores (QS) across quantiles. QS for $p=0.5$ corresponds to the mean absolute error (MAE). The benchmark shows actual values shaded in light grey, all other models are shown relative to the benchmark. The best performing model is indicated in dark grey shading.}
\end{tablenotes}
\end{threeparttable}
\end{scriptsize}
\end{center}
\caption{Quantile scores for UK inflation.} 
\label{tab:UK-QS}
\end{table*}

\begin{table*}[ht]
\begin{center}
\begin{scriptsize}
\begin{threeparttable}
\begin{tabular*}{\textwidth}{@{\extracolsep{\fill}} lrrrrrrrrrrrrrrr}
  \toprule
 & \multicolumn{15}{c}{\textbf{QS}} \\
 \cmidrule(l){2-16}
\multicolumn{1}{r}{$p=$} & \multicolumn{3}{c}{$0.05$} & \multicolumn{3}{c}{$0.1$} & \multicolumn{3}{c}{$0.5$ (MAE)} & \multicolumn{3}{c}{$0.9$} & \multicolumn{3}{c}{$0.95$} \\
 \cmidrule(l){2-4}\cmidrule(l){5-7}\cmidrule(l){8-10}\cmidrule(l){11-13}\cmidrule(l){14-16}
\multicolumn{1}{r}{$h=$} & 1 & 4 & 12 & 1 & 4 & 12 & 1 & 4 & 12 & 1 & 4 & 12 & 1 & 4 & 12 \\
  \midrule
UC-SV & \colBench$0.28$ & \colBench$0.37$ & \colBench$0.61$ & \colBench$0.44$ & \colBench$0.58$ & \colBench$0.82$ & \colBench$0.85$ & \colBench$1.03$ & \colBench$1.17$ & \colBench$0.39$ & \colBench$0.47$ & \colBench$0.50$ & \colBench$0.24$ & \colBench$0.30$ & \colBench$0.30$ \\ 
  UC-SVM & $1.00$ & $1.22$ & $0.97$ & \colBest$0.97$ & $1.10$ & $0.97$ & \colBest$0.97$ & $1.07$ & $1.06$ & \colBest$0.93$ & $1.00$ & $1.03$ & \colBest$0.93$ & $1.00$ & $1.01$ \\ 
  \midrule
  \multicolumn{16}{c}{\textbf{UCQR-TIS}} \\
  \midrule
   dhs & $1.21$ & $1.34$ & $1.15$ & $1.09$ & $1.17$ & $1.08$ & $1.04$ & $1.01$ & $1.00$ & $0.99$ & $1.02$ & $0.96$ & $1.03$ & $1.09$ & $1.02$ \\ 
  shs & $1.23$ & $1.37$ & $1.16$ & $1.10$ & $1.20$ & $1.09$ & $1.01$ & $1.04$ & $1.03$ & $0.98$ & $1.07$ & $1.00$ & $1.04$ & $1.14$ & $1.13$ \\ 
  iG & $1.30$ & $1.62$ & $1.39$ & $1.13$ & $1.27$ & $1.23$ & $0.98$ & $1.02$ & $1.03$ & $1.00$ & $1.14$ & $1.07$ & $1.06$ & $1.30$ & $1.32$ \\ 
  dhs-GP & $1.20$ & $1.34$ & $1.16$ & $1.11$ & $1.17$ & $1.07$ & $1.03$ & $1.02$ & $1.00$ & $1.00$ & $1.02$ & $0.95$ & $1.05$ & $1.09$ & $1.04$ \\ 
  shs-GP & $1.23$ & $1.38$ & $1.17$ & $1.10$ & $1.19$ & $1.09$ & $1.01$ & $1.04$ & $1.03$ & $0.99$ & $1.07$ & $1.00$ & $1.04$ & $1.14$ & $1.13$ \\ 
  iG-GP & $1.23$ & $1.37$ & $1.16$ & $1.11$ & $1.20$ & $1.09$ & $1.01$ & $1.03$ & $1.03$ & $0.99$ & $1.07$ & $1.01$ & $1.03$ & $1.13$ & $1.12$ \\ 
  dhs-GPt & $1.28$ & $1.59$ & $1.35$ & $1.13$ & $1.28$ & $1.25$ & $0.98$ & $1.02$ & $1.02$ & $1.00$ & $1.16$ & $1.09$ & $1.06$ & $1.28$ & $1.29$ \\ 
  shs-GPt & $1.34$ & $1.62$ & $1.43$ & $1.14$ & $1.28$ & $1.24$ & $0.98$ & $1.01$ & $1.03$ & $1.02$ & $1.15$ & $1.08$ & $1.10$ & $1.31$ & $1.28$ \\ 
  iG-GPt & $1.20$ & $1.34$ & $1.17$ & $1.10$ & $1.17$ & $1.06$ & $1.03$ & $1.02$ & $1.00$ & $1.00$ & $1.02$ & $0.95$ & $1.05$ & $1.08$ & $1.03$ \\ 
  \midrule
  \multicolumn{16}{c}{\textbf{UCQR-TVS}} \\
  \midrule
  dhs & $1.07$ & $0.91$ & $0.58$ & $1.10$ & $0.95$ & $0.71$ & $1.12$ & \colBest$0.95$ & $0.91$ & $1.09$ & $0.94$ & $0.93$ & $1.11$ & $0.94$ & $0.98$ \\ 
  shs & $1.05$ & \colBest$0.81$ & \colBest$0.47$ & $1.14$ & $0.98$ & $0.71$ & $1.12$ & $0.96$ & $0.92$ & $1.03$ & \colBest$0.89$ & $0.93$ & $1.12$ & $0.93$ & $1.06$ \\ 
  iG & $1.02$ & $0.85$ & $0.53$ & $1.05$ & $0.92$ & $0.72$ & $1.08$ & $0.98$ & $0.89$ & $1.04$ & $0.91$ & $0.91$ & $1.14$ & $0.97$ & $1.01$ \\ 
  dhs-GP & $1.09$ & $0.91$ & $0.57$ & $1.11$ & $0.98$ & $0.73$ & $1.11$ & $0.95$ & $0.91$ & $1.06$ & $0.93$ & $0.91$ & $1.11$ & $0.93$ & $0.98$ \\ 
  shs-GP & $1.09$ & $0.91$ & $0.59$ & $1.13$ & $0.99$ & $0.74$ & $1.12$ & $0.96$ & $0.91$ & $1.04$ & $0.92$ & $0.91$ & $1.08$ & $0.93$ & \colBest$0.96$ \\ 
  iG-GP & $1.10$ & $0.91$ & $0.59$ & $1.11$ & $0.96$ & $0.72$ & $1.11$ & $0.96$ & $0.91$ & $1.07$ & $0.93$ & $0.92$ & $1.08$ & $0.93$ & $0.96$ \\ 
  dhs-GPt & $1.02$ & $0.87$ & $0.53$ & $1.06$ & \colBest$0.90$ & \colBest$0.71$ & $1.08$ & $0.98$ & \colBest$0.89$ & $1.05$ & $0.93$ & $0.93$ & $1.13$ & $0.96$ & $1.00$ \\ 
  shs-GPt & $1.16$ & $1.17$ & $0.70$ & $1.19$ & $1.18$ & $0.85$ & $1.09$ & $0.98$ & $0.89$ & $0.98$ & $0.98$ & \colBest$0.86$ & $0.96$ & $1.00$ & $0.97$ \\ 
  iG-GPt & $1.02$ & $0.92$ & $0.55$ & $1.14$ & $1.00$ & $0.75$ & $1.11$ & $0.95$ & $0.91$ & $1.05$ & $0.91$ & $0.92$ & $1.13$ & \colBest$0.93$ & $1.00$ \\ 
   \bottomrule
\end{tabular*}
\begin{tablenotes}[para,flushleft]
\scriptsize{\textit{Notes}: For model abbreviations, see Section \ref{subsec:UCQR} and Tab. \ref{tab:models}. Quantile scores (QS) across quantiles. QS for $p=0.5$ corresponds to the mean absolute error (MAE). The benchmark shows actual values shaded in light grey, all other models are shown relative to the benchmark. The best performing model is indicated in dark grey shading.}
\end{tablenotes}
\end{threeparttable}
\end{scriptsize}
\end{center}
\caption{Quantile scores for EA inflation.} 
\label{tab:EA-QS}
\end{table*}

\clearpage
\begin{center}
{\Large\sffamily\textbf{Online Appendix}}
\end{center}

\setcounter{section}{0}
\setcounter{equation}{0}

\renewcommand\theequation{A.\arabic{equation}}
\renewcommand\thefigure{A.\arabic{figure}}
\renewcommand\thesection{A.\arabic{section}}
\section{Additional empirical results}
\subsection{In-sample results}
This Section provides several additional in-sample empirical results. I produce the same charts as in Fig. \ref{fig:insample} using an inverse Gamma prior (\texttt{iG}) on the state innovation variances rather than the dynamic horseshoe prior.

\begin{figure}[ht]
  \begin{subfigure}{\textwidth}
  \caption{US}\vspace*{-0.75em}
  \includegraphics[width=\textwidth]{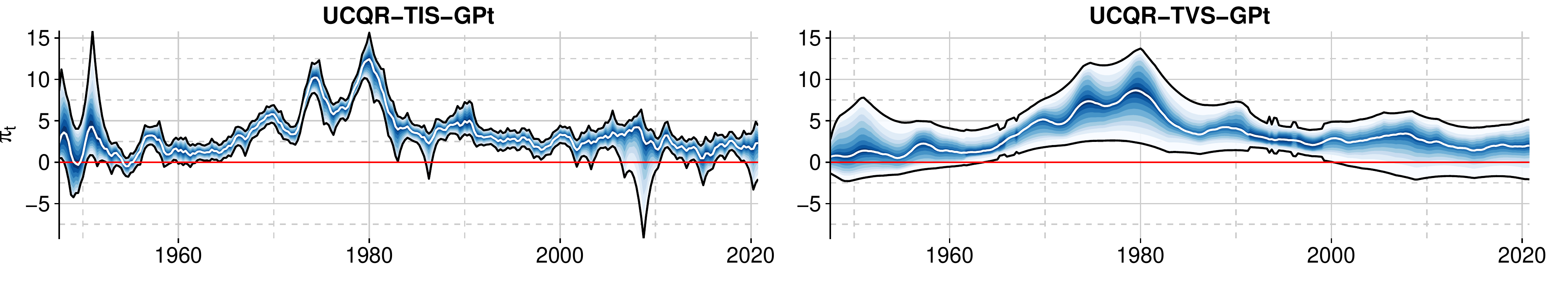}
  \end{subfigure}
  \begin{subfigure}{\textwidth}
  \caption{UK}\vspace*{-0.75em}
  \includegraphics[width=\textwidth]{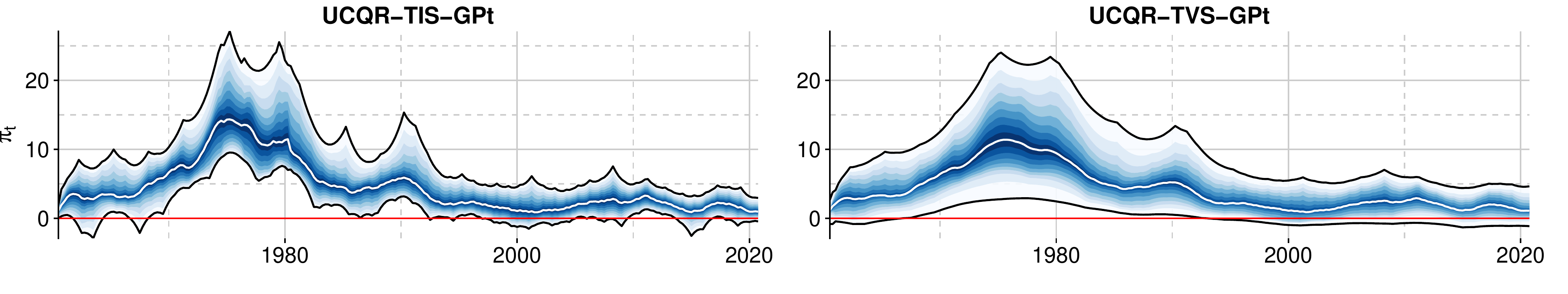}
  \end{subfigure}
  \begin{subfigure}{\textwidth}
  \caption{EA}\vspace*{-0.75em}
  \includegraphics[width=\textwidth]{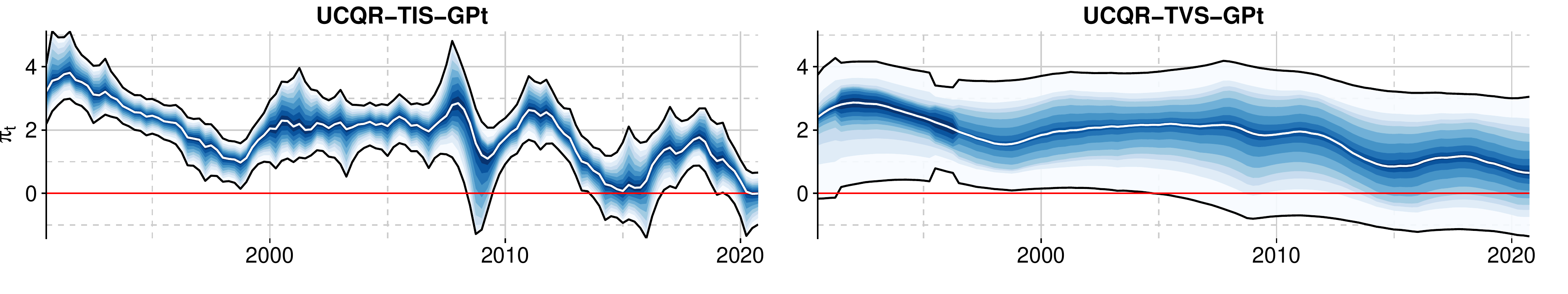}
  \end{subfigure}
  \caption{Unobserved components for variants of UCQR with inverse Gamma prior.}\label{fig:insample_app}\vspace*{-1em}
  \caption*{\footnotesize{\textit{Notes:} Unobserved component quantile regression (UCQR) with time-invariant (TIS) or time-varying (TVS) scale parameter, adjusted using a Gaussian process (GPt) regression with time-varying bandwidth. The red line marks zero. The solid black lines are the posterior mean of the $5$th and $95$th quantiles, the solid white line is the $50$th percentile (median). The blue shaded areas cover quantile pairs (e.g., $10$th to $90$th percentile) in increments of five.}}
\end{figure}

\end{appendices}
\end{document}